\documentclass[journal]{IEEEtran}

\usepackage{amsmath,amsfonts,amssymb}
\usepackage{graphicx}
\usepackage{bm} 
\usepackage{textcomp} 
\usepackage{xcolor}
\usepackage{cite}
\usepackage{hyperref}
\newcommand{\ket}[1]{\ensuremath{|#1\rangle}}
\newcommand{\bra}[1]{\ensuremath{\langle#1|}}

\newcommand{\subrm}[1]{\ensuremath{_{\text{#1}}}}

\begin{document}
\title{\huge Rydberg Atomic Quantum Receivers for Wireless
Communications: Two-Color vs. Three-Color Excitation}

\author{Jian Xiao, Tierui Gong, Ji Wang, Erry Gunawan,
        and Chau Yuen,~\IEEEmembership{Fellow,~IEEE}
\thanks{

Jian Xiao is with the Department of Electronics and Information Engineering, Central China Normal University, Wuhan 430079, China, and also with the School of Electrical and Electronics Engineering, Nanyang Technological University, Singapore 639798. (e-mail: jianx@mails.ccnu.edu.cn).

Ji Wang is with the Department of Electronics and Information Engineering, College of Physical Science and Technology, Central China Normal University, and is also with Academy of Frontier Interdisciplinary Research, Central China Normal University, Wuhan 430079, China (e-mail: jiwang@ccnu.edu.cn).

Tierui Gong, Erry Gunawan and Chau Yuen are with the School of Electrical and Electronics Engineering, Nanyang Technological University, Singapore 639798.
(e-mail: tierui.gong@ntu.edu.sg; egunawan@ntu.edu.sg; chau.yuen@ntu.edu.sg).

}}

\maketitle

\begin{abstract}

An efficient three-color {(3C)} laser excitation-based Rydberg atomic quantum receiver (RAQR) architecture is investigated for wireless communications, utilizing a five-level {(5L)} electronic transition mechanism.
Specifically, the conventional two-color {(2C)} RAQR with the four-level {(4L)} excitation faces three fundamental obstacles: 1) high cost and engineering challenges due to the reliance on unstable {short-wavelength} lasers; 2) a fundamental sensitivity limit in thermal atoms caused by residual Doppler broadening; and 3) the inability to detect low-frequency bands due to the energy-level constraint of two-photon resonance. To address these challenges, this paper analyzes a {3C5L-RAQR} architecture with all-red/infrared lasers, which not only solves the engineering cost issues but also enables effective Doppler cancellation and low-frequency detection by exploiting the three-photon resonance. Bridging atomic physics and communication theory, an end-to-end equivalent baseband signal model is derived. Furthermore, the performance of different RAQR architectures is evaluated in terms of sensitivity, achievable {rate} and spectrum access range. Moreover, we provide an exact numerical solution for practical RAQRs by employing the Liouvillian superoperator formalism. Numerical results demonstrate that the exhibited 3C5L-RAQR achieves superior sensitivity compared to the conventional {2C4L-RAQR} and a classical antenna-based radio frequency receiver {for weak-signal detection}. Finally, the inherent sensitivity-rate trade-off is revealed, showing that the 3C5L-RAQR is more suitable for deployment in power-limited communication scenarios demanding broad spectrum access.

\end{abstract}


\section{Introduction}
\label{sec:introduction}
\IEEEPARstart{R}{ydberg} atomic sensors have been extensively investigated for decades as highly sensitive detection tools, leveraging the giant electric dipole moments of Rydberg atoms \cite{saffman2010quantum, yuan2023quantum}. More recently, a significant research trend has emerged that seeks to extend the capabilities of these atomic systems from pure sensing to the domain of wireless communications \cite{meyer2018digital, 9069423}. While the Rydberg atom configuration exhibits enhanced sensitivity via coherent control, these gains are inevitably accompanied by an inherent trade-off that follows the no free lunch theorem for any scientific research \cite{585893}. Specifically, the heightened complexity of the atom resonance alignment by expensive and large laser equipment and the narrower spectral windows suggest that the quantum advantage may be constrained to specific operational regimes \cite{somaweera2025rydberg}.

\subsection{Background of Rydberg Atomic Quantum Receivers}

Historically, the application of Rydberg atoms to the radio frequency (RF) domain progressed through two primary stages. The first phase, primarily driven by the physics and quantum optics communities, focused on fundamental experimental demonstrations of RF signal detection with Rydberg atoms \cite{sedlacek2012microwave}. During this period, significant experimental progress has been made in demonstrating the capabilities of RAQRs, including digital communications \cite{meyer2018digital} using electromagnetically induced transparency (EIT) effect, signal reception under various modulations \cite{ref_holloway_psk}, and multi-band detection by utilizing different Rydberg transitions \cite{ref_meyer_multiband}.

The second stage, which rose to prominence from 2024 to the present, marks a shift toward a telecommunications-oriented perspective. Researchers started leveraging Rydberg physical models for various communication scenarios, e.g., fundamental performance analysis \cite{ref_gong_overview, chen-Rydberg1}, multiple-input multiple-output (MIMO) communications \cite{ref_cui_mimo, 10972179}, and wireless sensing \cite{11301641, 11429587, chen-Rydberg2}. In this stage, this receiver architecture is termed the Rydberg atomic quantum receiver (RAQR) \cite{10972179} or Rydberg atomic receiver (RARE) \cite{11429587}. Unlike the experimental nature of the first phase, this recent wave of research is predominantly theoretical, relying on simulations to verify concepts. As such, these system-level applications are still in their infancy and require further validation.

\subsection{Motivations for New RAQR Architectures}

Currently, the research landscape of RAQR-aided wireless systems is predominantly focused on the conventional two-color (2C) laser architecture \cite{ref_cui_mimo, 10972179, 11301641, 11429587}, employing a ladder-type excitation scheme to realize the four-level (4L) electronic transition of Rydberg atoms. The {2C4L-RAQR} utilizes a probe laser and a coupling laser to drive the atoms into a coherent dark state via two-photon resonance. However, despite the established efficacy of the conventional 2C4L-RAQR, it is constrained by three fundamental limitations:

\begin{enumerate}
    \item \textbf{The Engineering and Cost Problem:} The reliance on a high-power short-wavelength laser, e.g., the wavelength of coupling laser $\lambda\subrm{c} \approx 480-510$ nm \cite{ref_thaicharoen_all_ir}, is a major engineering barrier. The {blue/green} laser is notoriously expensive, physically bulky, and high-power \cite{zuo2025high}, which makes the standard 2C4L-RAQR impractical for low-cost, compact or field-deployable units.

\item \textbf{The Fundamental Sensitivity Limit:} In general, the sensitivity of RAQR is inversely proportional to the linewidth of EIT. In a room-temperature thermal vapor cell, the 4L transition scheme with 2C lasers suffers from a large wavelength mismatch between the infrared probe and the {blue/green} coupling lasers. The residual Doppler broadening creates a fundamental floor for the EIT linewidth and masks the atomic response to weak signals, reducing the receiver sensitivity \cite{ref_prajapati_sensitivity}.

\item \textbf{The Frequency Access Limit:} In the typical 2C4L excitation scheme, atoms are driven to a Rydberg $D$-state with orbital angular momentum $l=2$ \cite[Table~1]{10972179}. Resonant detection from these $D$-states is physically constrained by the large energy gaps to the adjacent allowed states, restricting the detectable frequencies to the high-GHz or THz range \cite{prajapati2024high}\footnote{For detecting strategic low-frequency bands, the conventional transition scheme in the 2C4L-RAQR can use inefficient off-resonant mechanisms, e.g., alternating current (AC) Stark shift \cite{song2024continuous, hu2022continuously}, but inherently limits sensitivity. While direct current (DC) Stark tuning can be used to force a resonance \cite{Simons2021Continuous, ouyang2023continuous}, this alternative method introduces significant line-broadening noise and technical complexity from the strong external DC field.}. The high-sensitivity detection relying on natural resonance of Rydberg atoms is fundamentally inaccessible to the 2C4L-RAQR in the low-frequency band, e.g., very-high frequency (VHF) to ultra-high frequency (UHF) bands.
\end{enumerate}

\begin{figure}[t]
	\centerline{\includegraphics[width=2.8in]{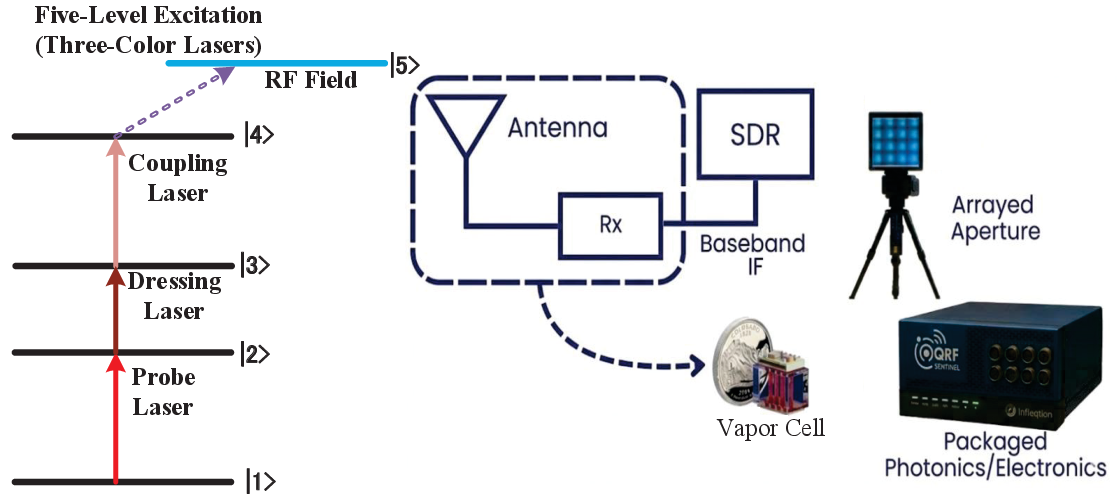}}
	 \caption{Proof-of-concept product of 3C5L-RAQR from Infleqtion Inc. \cite{ref_infleqtion_sqywire}. 
	 }
	\label{Three-photon RAQR}
\end{figure}

To address the above limitations, recent physics experiments explored three-color (3C) RAQR architecture, i.e., utilizing three lasers with distinct wavelengths to construct the five-level (5L) excitation of Rydberg atoms. The authors of \cite{ref_thaicharoen_all_ir} demonstrated an all-infrared 3C scheme to achieve EIT, proving the feasibility of low-cost diode lasers. Furthermore, the authors of \cite{ref_prajapati_sensitivity} demonstrated that 3C5L electrometry yields a narrower linewidth and higher sensitivity compared to the 2C4L architecture due to superior Doppler cancellation. In \cite{brown2023very}, the resonant detection of RF electric fields from VHF to UHF bands was implemented by using {3C5L-RAQR} to construct the high orbital angular momentum Rydberg states. {Moreover, Fig.~\ref{Three-photon RAQR} presents recent industrial breakthroughs, exemplified by the Infleqtion Sqywire \cite{ref_infleqtion_sqywire}, which implements three-color RAQR architectures in compact commercial modules. }

While the theoretical physical foundations of 3C5L-RAQR have shown great potential as a new architecture, these existing works are predominantly focused on static field electrometry or spectroscopy. However,
the communication-centric modeling and analysis for 3C5L-RAQR assisted wireless communications remain largely unexplored.

\subsection{Contributions}
Against the aforementioned background, in this paper, we investigate the 3C5L-RAQR architecture for wireless communications. Our main contributions are summarized as follows:

\begin{itemize}
    \item We establish the physical model for the 3C5L excitation scheme with low-cost all-red/infrared lasers and identify the residual Doppler broadening as the fundamental sensitivity bottleneck in thermal vapor cells for practical RAQRs. Furthermore, we derive an end-to-end equivalent baseband signal model that maps the physical atom-field interaction to a standard communication model. 

    \item We formulate the scaling laws of SNR for 2C4L- and 3C5L-RAQRs in different quantum noise regimes under the weak probe field. Then, we reveal that the achievable {rate} of the RAQR is restricted by the limited instantaneous bandwidth that is related to the relaxation time of Rydberg atoms. We also prove that the 3C5L-RAQR serves as a unified platform capable of accessing strategic UHF/VHF bands via high-angular-momentum transitions.
    
        \item We propose an efficient numerical modeling of open quantum dynamics for practical RAQRs, which surpasses the classic assumption of the weak probe field to achieve the superior receiver performance. By employing the Liouvillian superoperator formalism, we provide the exact numerical solution for atomic coherence and {baseband modulation bandwidth of the superheterodyne 3C5L-RAQR architecture.}
    
    \item
    We demonstrate through numerical results that the exhibited 3C5L-RAQR with thermal robustness outperforms the conventional 2C4L-RAQR and the classical RF receiver in terms of signal-to-noise ratio (SNR) gain of {weak signal detection}, and block error rate (BLER) performance. Due to limited instantaneous bandwidth, the RAQR is more suitable for deployment in power-limited and spectrum-agile communication scenarios, while the classical RF receiver retains an advantage in high-speed communications. 
    
\end{itemize}


\emph{Notations:} $\ket{i}$ denotes the $i$-th atomic state and $\bra{i}$ denotes its corresponding bra vector. $\bra{i}\hat{\rho}\ket{j}$ represents the matrix element of density operator $\hat{\rho}$ between state $\ket{i}$ and $\ket{j}$. The dot notation in $\dot{\rho}$ represents the first time derivative $d\rho/dt$. 
$\ket{nL_J}$ denotes an atomic state, where \(n\) is the principal quantum number, \(L\in\{S,P,D,F,\ldots\}\) is the spectroscopic orbital label corresponding to the orbital angular momentum quantum number \(l=0,1,2,3,\ldots\), respectively, and \(J\) is the total angular momentum quantum number.
Operators $\Re(\cdot)$, $\Im(\cdot)$ and $\text{arg}(\cdot)$ denote the real, imaginary and phase components of the complex-value elements. Operator $(\cdot)^{*}$ denotes the conjugate of complex-value elements. Operators $(\cdot)^{-1}$ and $\det(\cdot)$ denote the inverse and determinant of a matrix, respectively. Operator $\text{Tr}(\cdot)$ denotes the trace of a matrix. Symbols $q_0$ and $a_0$ denote elementary charge and Bohr radius, respectively.

\section{Physical Architectures and Signal Models}
\label{sec:system_models}
In this section, we first establish the physical models for different RAQR architectures. Then, we derive the communication baseband signal model of the exhibited 3C5L-RAQR.

\begin{figure}[t]
	\centerline{\includegraphics[width=3.0in]{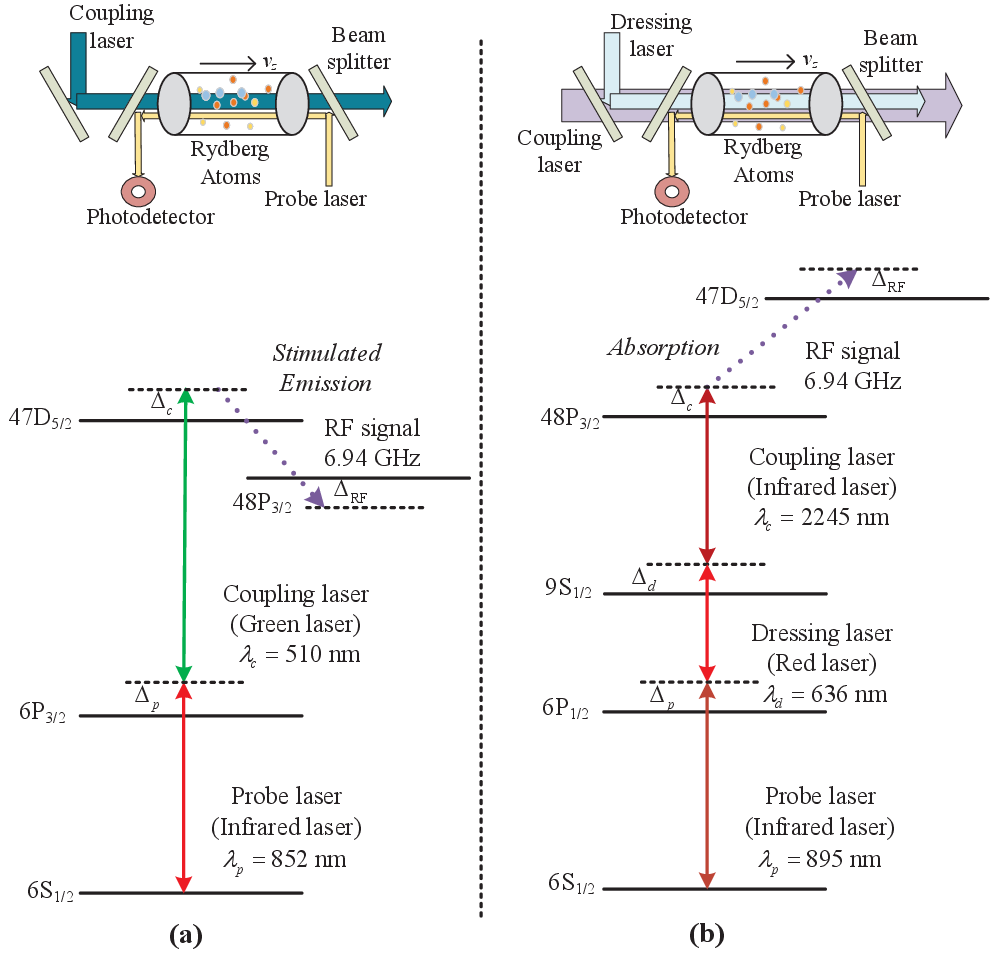}}
	 \caption{Atomic energy level schemes for Cs atoms \cite{ref_prajapati_sensitivity}. (a) The conventional 2C4L-RAQR with electronic transition path $6S \to 6P \to nD$, which is driven by an infrared probe laser with wavelength $\lambda\subrm{p} = 852$ nm and a high-power {green} coupling laser with wavelength $\lambda\subrm{c} = 510$ nm. (b) The exhibited 3C5L-RAQR with transition path $6S \to 6P \to 9S \to n'P$, which uses low-cost diode lasers, i.e., an infrared probe laser with wavelength $\lambda\subrm{p}=895$ nm, a red dressing laser with wavelength $\lambda\subrm{d}=636$ nm, and an infrared coupling laser with wavelength $\lambda\subrm{c}= 2245$ nm. Note that considering the same incident RF signal, the 2C4L-RAQR carries out the downward electronic transitions based on the stimulated emission mechanism \cite{9087968}, i.e., $\ket{47D_{5/2}} \to \ket{48P_{3/2}}$, while the 3C5L-RAQR achieves the upward transitions via energy absorption, i.e., $\ket{48P_{3/2}} \to \ket{47D_{5/2}}$.}
	\label{fig:level_schemes}
\end{figure}

\subsection{Conventional 2C4L-RAQR Architectures}
\label{ssec:4L_model}
{In this section, we review the physical architectures of the conventional 2C4L-RAQR to provide a comparative baseline.}
\subsubsection{System Architecture} As shown in Fig. \ref{fig:level_schemes}(a), the standard 2C4L-RAQR operates on a 4L electronic transition scheme. A common example in Cesium (Cs) atoms is $\ket{1} = \ket{6S_{1/2}} \to \ket{2} = \ket{6P_{3/2}} \to \ket{3} = \ket{nD_J}$. The microwave signal couples $\ket{3}$ to an adjacent state $\ket{4}$.
This excitation pathway requires an infrared probe laser with wavelength $\lambda\subrm{p}$ and a {green} coupling laser with wavelength $\lambda\subrm{c}$. This reliance on a high-power {green} laser $\lambda\subrm{c} = 510$ nm constitutes the major engineering problem\footnote{{In typical 2C4L-RAQR systems, the required wavelength for the coupling laser inherently depends on the specific alkali metal employed in the vapor cell. Specifically, the foundational experiments with Rubidium (Rb) atoms generally require a coupling laser at approximately 480 nm \cite{Holloway2014Broadband}, falling within the blue region of the visible spectrum. This work considers Cs atoms-based RAQRs, which dictate a transition pathway requiring a ~510 nm {green} coupling laser \cite{ref_prajapati_sensitivity}. Note that both blue/green lasers share the same fundamental engineering bottleneck, i.e., generating high-power and narrow-linewidth light at either of these wavelengths heavily relies on highly sensitive and expensive second-harmonic generation modules \cite{zuo2025high}.}}, which makes the standard 2C4L-RAQR impractical for low-cost, compact, or field-deployable units.

\subsubsection{Physical Modeling}
The coherent dynamics of the 2C4L electronic transition are driven by a probe laser, a coupling laser, and an RF signal field. The evolution is governed by the Lindblad master equation \cite{berman2011principles, holloway2017electric}, which is given by
\begin{equation}
\frac{d\hat{\rho}}{dt} = -\frac{i}{\hbar}[\hat{H}_{\text{4L}}, \hat{\rho}] + \mathcal{L}(\hat{\rho}),
\label{eq:rx_master_eq}
\end{equation}
where $\hat{\rho}$ is the atomic density matrix and $\hbar$ denotes the reduced Planck constant. $\hat{H}_{\text{4L}}$ denotes the Hamiltonian for the 2C4L-RAQR system. $\mathcal{L}(\hat{\rho})$ describes all spontaneous decay and dephasing processes.

To find the steady-state coherence ${\rho}_{21}=\bra{2}\hat{\rho}\ket{1}$, we solve $\dot{\hat{\rho}}=\frac{d\hat{\rho}}{dt} = 0$. By utilizing the specific assumptions and approximations, different analytical solutions of ${\rho}_{21}$ have been derived \cite{ref_gong_overview, ref_doppler_mismatch, holloway2017electric}. Considering the classic weak-probe approximation\footnote{From the perspective of sensitivity optimization, the classic weak-probe approximation in the RAQR is suboptimal \cite{meyer2021optimal}, particularly for the practical thermal atoms with severe Doppler broadening. In this section, to gain theoretical and explainable insights into the fundamental sensitivity bottlenecks for the subsequent performance analysis with an analytical form, we first present the atomic coherence under the ideal weak-probe approximation, while we provide the exact numerical modeling of ${\rho}_{21}$ for general probe fields in Section IV.}, the steady-state coherence ${\rho}_{21}^\text{4L}$ for 2C4L-RAQR can be expressed as \cite{ref_doppler_mismatch}
\begin{equation}
{\rho}_{21}^{{\text{4L}}} = \frac{i\Omega\subrm{p}^{{\text{4L}}}/2}{(i\Delta_2^{{\text{4L}}} - \Gamma_{21}) + \frac{|\Omega\subrm{c}^{{\text{4L}}}|^2/4}{ (i\Delta_3^{{\text{4L}}} - \Gamma_{31}) + \frac{|\Omega\subrm{RF}|^2/4}{i\Delta_4^{{\text{4L}}} - \Gamma_{41}} }},
\end{equation}
where $\Omega\subrm{p}^{{\text{4L}}}$, $\Omega\subrm{c}^{{\text{4L}}}$, and $\Omega\subrm{RF}$ are the Rabi frequencies of the probe, coupling, and RF fields for the 2C4L-RAQR, respectively. 
$\Delta_2^{{\text{4L}}} = \Delta\subrm{p}^{{\text{4L}}}$, $\Delta_3^{{\text{4L}}} = \Delta\subrm{p}^{{\text{4L}}} + \Delta\subrm{c}^{{\text{4L}}}$, and $\Delta_4 = \Delta_3^{{\text{4L}}} + \Delta\subrm{RF}$. Here, $\Delta\subrm{p}^{{\text{4L}}}$, $\Delta\subrm{c}^{{\text{4L}}}$, and $\Delta\subrm{RF}$ are the corresponding frequency detunings from their atomic resonances. $\Gamma_{n1}, n \in\{2, \ldots, 5\}$ are the coherence decay rates between states $\ket{n}$ and $\ket{1}$. Each $\Gamma_{n1}$ is the sum of all processes that destroy the phase coherence between the two states. If collisional dephasing is negligible, we have $\Gamma_{n1} = \frac{1}{2}(\Gamma_n + \Gamma_1)$ \cite{ref_doppler_mismatch}, where $\Gamma_n$ and $\Gamma_1$ are the natural decay rates of the excited state $|n\rangle$ and the ground state, respectively. $\Gamma_1$ is usually set to 0 as the ground state is stable, so the relation simplifies to $\Gamma_{n1} = \frac{\Gamma_n}{2}$. Note that in the EIT linewidth modeling of Section III-A, the detailed components for total dephasing rate are investigated.

The optical response of the receiver is determined by the bulk atomic susceptibility $\chi$, which can be expressed as \cite{holloway2017electric}
\begin{equation}
\chi = \frac{2N_a |\mu_{12}|^2}{\epsilon_0 \hbar \Omega\subrm{p}} {\rho}_{21},
\label{eq:rx_susceptibility}
\end{equation}
where $N_a$ is the effective density of atoms populating the ground state $|1\rangle$, $\mu_{12}$ is the dipole moment of the $\ket{1} \to \ket{2}$ transition, and $\epsilon_0$ is the vacuum permittivity.

\subsubsection{Physical Limitation}

\begin{figure}[t]
	\centerline{\includegraphics[width=2.8in]{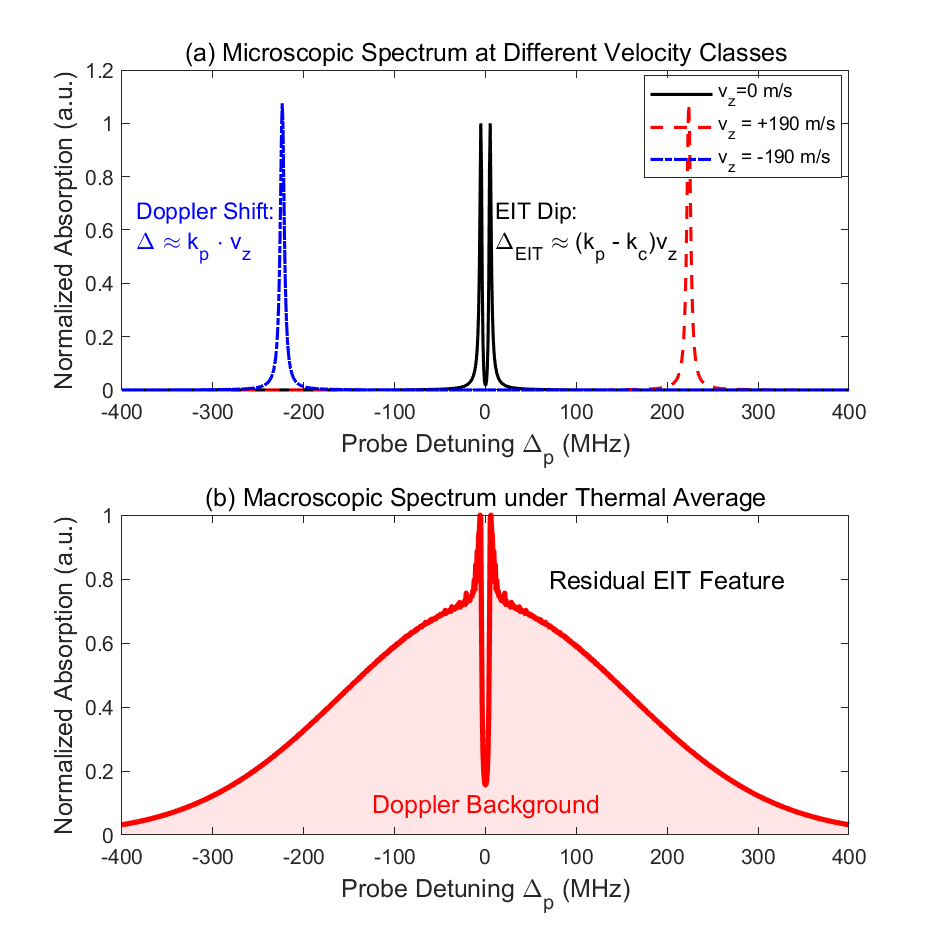}}
\caption{EIT absorption spectrum $\Im(\hat{\rho}_{21}^{{\text{4L}}})$ vs. probe detuning $\Delta\subrm{p}$.}
\label{fig:4L_spectrum1}
\end{figure}

The EIT mechanism in the 2C4L-RAQR relies on precise ladder-type transition resonance between the probe and coupling laser fields, i.e., $\Delta\subrm{p}^{{\text{4L}}} + \Delta\subrm{c}^{{\text{4L}}}= 0$. In a thermal vapor cell, an atom moving with velocity $\vec{v}$ perceives the laser frequencies shifted by the Doppler effect $\Delta_j(\vec{v}) = \Delta_j - \vec{k}_j  \vec{v}$, where $\vec{k}_j$ is the wavevector of the $j$-th laser field.
To minimize the atomic Doppler mismatch, the probe and coupling beams are configured to be counter-propagating. We define the propagation direction of the probe beam as the positive $z$-axis and $v_z$ is the longitudinal velocity component of thermal atom motion, such that $\vec{k}\subrm{p} = k\subrm{p}\hat{z}$ and $\vec{k}\subrm{c} = -k\subrm{c}\hat{z}$, where $k_j = 2\pi/\lambda_j$ denotes the wavenumber. Consequently, for an atom moving with a velocity vector $\vec{v} \approx v_z\hat{z}$ along the optical axis at a vapor-cell temperature $T_\text{atom}$, the effective detuning in the 2C4L-RAQR is given by \cite{mohapatra2007coherent, ref_doppler_mismatch}
\begin{align}
\Delta\subrm{EIT}^{{\text{4L}}} &= (\Delta\subrm{p}^{{\text{4L}}} - \vec{k}\subrm{p}\cdot\vec{v}) + (\Delta\subrm{c}^{{\text{4L}}} - \vec{k}\subrm{c}\cdot\vec{v}) \nonumber \\
&= (\Delta\subrm{p}^{{\text{4L}}}+\Delta\subrm{c}^{{\text{4L}}}) + (k\subrm{c} - k\subrm{p})v_z.
\label{ff}
\end{align}



\begin{figure}[t]
	\centerline{\includegraphics[width=2.8in]{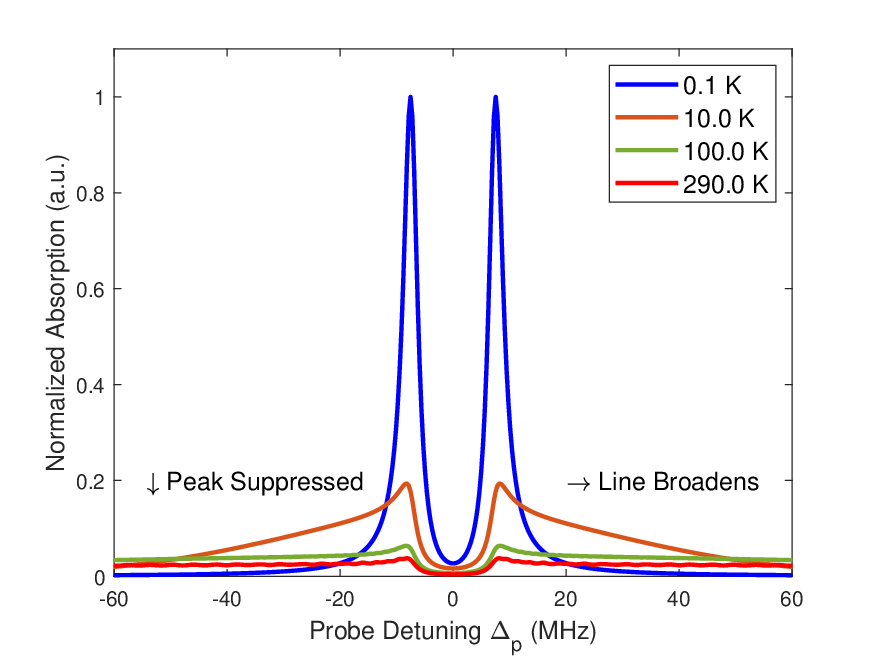}}
\caption{EIT absorption spectrum $\Im(\hat{\rho}_{21}^{{\text{4L}}})$ vs. probe detuning $\Delta\subrm{p}$ for different vapor-cell temperatures $T_\text{atom}$.}
\label{fig:4L_spectrum2}
\end{figure}

Fig. \ref{fig:4L_spectrum1} illustrates the microscopic origin of the sensitivity limit in thermal vapor cells. (a) Individual velocity classes of atoms experience different Doppler shifts, effectively smearing the resonance across the frequency domain. (b) The macroscopic response under Doppler-averaging, obtained by integrating over the Maxwell-Boltzmann distribution \cite{meyer2021optimal, berman2011principles, ref_prajapati_sensitivity, holloway2017electric}, reveals that the sharp EIT feature is severely broadened into a wide Doppler background\footnote{{An ideal laser setup in 3C5L-RAQR is to construct a zero Doppler frequency shift, i.e., $|k\subrm{p} - k\subrm{d} + k\subrm{c} |=0$, which can achieve the perfect Doppler cancellation. In this work, we comprehensively consider the practical engineering implementation and performance trade-offs. Hence, we refer to the three-photon configuration in \cite{ref_prajapati_sensitivity} that still has very little Doppler frequency shift due to atom motion. Given the laser setups in Fig. \ref{fig:level_schemes}, for the 2C4L-RAQR, the large effective wavevector mismatch $k_\text{eff}= |k\subrm{p} - k\subrm{c} | \approx 4.9$ rad/$\mu$m results in a large frequency shift, while the 3C5L-RAQR exhibits a near-zero slope $k_\text{eff} = |k\subrm{p} - k\subrm{d} + k\subrm{c} | \approx 0. 06$ rad/$\mu$m.}}. 
Fig. \ref{fig:4L_spectrum2} further presents the impact of thermal field on the EIT linewidth. As the temperature $T_\text{atom}$ of the vapor cell increases from quasi-cold state 0.1 K to room-temperature state 290 K, the thermal velocity of the atoms increases, leading to severe Doppler broadening, where the EIT window widens significantly. 

Due to the large wavelength difference between the probe and coupling lasers, the wavevectors do not match. Even with counter-propagating beams, the cancellation is imperfect, i.e., $|k_p - k_c| \gg 0$, leading to a partial cancellation of the Doppler broadening effect. 
According to \cite{ref_prajapati_sensitivity}, the uncancelled residual Doppler linewidth in the 2C4L-RAQR can be expressed as
\begin{equation}
\Gamma_{\text{Res}}^{{\text{4L}}} = \frac{\Gamma_{2}}{|k_p|} |k\subrm{p} - k\subrm{c}|,
\label{eq:gamma_res_4L}
\end{equation}
where $\Gamma_2$ is the decay rate of the first excited state $\ket{2}$. This calculation yields a large residual linewidth $\Gamma_{\text{Res}}^{{\text{4L}}}$ on the order of several MHz. 

The fundamental sensitivity of the RAQR is limited by the EIT linewidth $\Gamma_{\text{EIT}}$, where a narrower $\Gamma_{\text{EIT}}$ generally results in a higher sensitivity\footnote{{Note that the optical Doppler broadening caused by thermal atom motion is different from the RF Doppler spread caused by wireless transceiver mobility. Considering a transmitter moving at $v_\mathrm{m}$ introduces a maximum RF Doppler shift $f_d = v_\mathrm{m} f_c / c$ at carrier frequency $f_c$, where $c$ denotes the speed of light. In the RAQR signal model, this translates to a dynamic detuning of the incident RF field, i.e., $\Delta_\mathrm{RF}' = \Delta_\mathrm{RF} + 2\pi f_d$. Since this RF Doppler shift is orders of magnitude smaller than the residual EIT linewidth $\Gamma_\mathrm{EIT}$, e.g., given $v_\mathrm{m} = 30\text{ m/s}$ and $f_c = 6.94\text{ GHz}$ which yields $f_d \approx 694\text{ Hz} \ll \Gamma_{\text{EIT}}$, the impinging signal remains firmly within the linear transconductance region of the atomic resonance against typical RF Doppler spreads.}}. Since the homogeneous broadening terms, e.g., natural linewidth, transit time, are typically on the order of kHz, the total EIT linewidth $\Gamma_{\text{EIT}}^{\text{4L}}$ for the 2C4L-RAQR is dominated by this residual Doppler mismatch $\Gamma_{\text{Res}}^{\text{4L}}$ \cite{ref_prajapati_sensitivity}. 
This broad MHz-scale linewidth flattens the dispersion curve, masking the atomic response to weak RF signals and creating a fundamental sensitivity bottleneck.
Moreover, the $S \to P \to D$ excitation pathway populates $D$-states, i.e., orbital angular momentum $l=2$. Allowed RF transitions from $D$-states, e.g., $nD \to n'P$, have large energy gaps, corresponding to high frequencies, e.g., GHz/THz. However, the 2C4L-RAQR architecture is physically incapable of detecting strategic low-frequency UHF (300 MHz$\sim$3 GHz) or VHF (30 MHz$\sim$300 MHz) bands.


\subsection{{Exhibited 3C5L-RAQR for Doppler-Free Excitation}}
\label{ssec:5L_model}

To overcome both the engineering and sensitivity limits inherent in the conventional 2C4L-RAQR configuration, we employ the Doppler-free 3C5L-RAQR excitation scheme recently demonstrated in physics experiments \cite{ref_prajapati_sensitivity, glick2025doppler}, as shown in Fig. \ref{fig:level_schemes}(b). In particular, we quantify the residual Doppler linewidths for both RAQR architectures and evaluate their macroscopic spectral responses.

\subsubsection{System Architecture} The excitation path is
$\ket{1} = \ket{6S_{1/2}} \xrightarrow{\Omega\subrm{p}^{\text{5L}}} \ket{2} = \ket{6P_{1/2}} \xrightarrow{\Omega\subrm{d}^{\text{5L}}} \ket{3} = \ket{9S_{1/2}} \xrightarrow{\Omega\subrm{c}^{\text{5L}}} \ket{4} = \ket{n'P_{J'}}$, where ${\Omega\subrm{p}^{\text{5L}}}$, ${\Omega\subrm{d}^{\text{5L}}}$, and ${\Omega\subrm{c}^{\text{5L}}}$ are the Rabi frequencies of the probe, dressing, and coupling fields for the 3C5L-RAQR, respectively.
The three fields are red/infrared lasers, which excite state $\ket{4}$ via three stable and low-cost diode lasers. 

\subsubsection{Physical Modeling} The dynamics in 3C5L-RAQR are governed by the master equation using the 5L Hamiltonian:
\begin{equation}
\hat{H}_{\text{5L}} = \frac{\hbar}{2}
\begin{pmatrix}
0 & \Omega\subrm{p}^{\text{5L}} & 0 & 0 & 0 \\
(\Omega\subrm{p}^{\text{5L}})^* & -2\Delta_2^{{\text{5L}}} & \Omega\subrm{d}^{\text{5L}} & 0 & 0 \\
0 & (\Omega\subrm{d}^{\text{5L}})^* & -2\Delta_3^{{\text{5L}}} & \Omega\subrm{c}^{\text{5L}} & 0 \\
0 & 0 & (\Omega\subrm{c}^{\text{5L}})^* & -2\Delta_4^{{\text{5L}}} & \Omega\subrm{RF} \\
0 & 0 & 0 & (\Omega_{\text{RF}})^* & -2\Delta_5^{{\text{5L}}}
\end{pmatrix},
\label{eq:rx_hamiltonian_5level}
\end{equation}
where $\Delta_2^{{\text{5L}}} = \Delta\subrm{p}^{{\text{5L}}}$, $\Delta_3^{{\text{5L}}} = \Delta\subrm{p}^{{\text{5L}}} + \Delta\subrm{d}^{{\text{5L}}}$, $\Delta_4^{{\text{5L}}} = \Delta_3^{{\text{5L}}} + \Delta\subrm{c}^{{\text{5L}}}$, $\Delta_5^{{\text{5L}}} = \Delta_4^{{\text{5L}}} + \Delta\subrm{RF}$. Here, $\Omega\subrm{d}^{{\text{5L}}}$ and $\Delta\subrm{d}^{{\text{5L}}}$ denote the newly introduced Rabi frequency and the corresponding frequency detuning of the dressing laser, respectively.

\textbf{Lemma 1:} Under the weak probe approximation \cite{ref_mondal_theory}, i.e., ${\rho}^\text{5L}_{11} \approx 1$ and ${\rho}^\text{5L}_{ii \ne 1} \approx 0$, the steady-state coherence ${\rho}_{21}^\text{5L}$ for the 3C5L-RAQR can be expressed as
\begin{equation}
{\rho}_{21}^{{\text{5L}}} = \frac{i\Omega\subrm{p}^{{\text{5L}}}/2}{(i\Delta_2 - \Gamma_{21}) + \frac{|\Omega\subrm{d}^{{\text{5L}}}|^2/4}{ (i\Delta_3 - \Gamma_{31}) + \frac{|\Omega\subrm{c}^{{\text{5L}}}|^2/4}{ (i\Delta_4 - \Gamma_{41}) + \frac{|\Omega\subrm{RF}|^2/4}{i\Delta_5 - \Gamma_{51}} } }}.
\label{eq:rho21_5level_continued_fraction}
\end{equation}
\textit{Proof:} Please refer to Appendix \ref{app:derivation_5level} for more details. \hfill $\blacksquare$

\subsubsection{Physical Advantages}
In the 3C5L-RAQR system, the EIT condition depends on three-photon resonance, i.e., $\Delta\subrm{p}^{{\text{5L}}} + \Delta\subrm{d}^{{\text{5L}}} + \Delta\subrm{c}^{{\text{5L}}} = 0$. We choose a counter-propagating geometry, e.g., $\vec{k}\subrm{p}$ and $\vec{k}\subrm{c}$ co-propagating, whereas $\vec{k}\subrm{d}$ is counter-propagating. The total velocity-dependent detuning becomes
\begin{equation}
\Delta\subrm{EIT}^{{\text{5L}}} = (\Delta\subrm{p}^{{\text{5L}}}+\Delta\subrm{c}^{{\text{5L}}}+\Delta\subrm{d}^{{\text{5L}}}) + (k\subrm{p} - k\subrm{d} + k\subrm{c}) v_z.
\end{equation}

Unlike the previous 4L excitation scheme dominated by a $k\subrm{p} - k\subrm{c}$ mismatch, the entire sum nearly vanishes. The residual Doppler broadening is given by
\begin{equation}
\Gamma_{\text{Res}}^{{\text{5L}}} = \frac{\Gamma_{2}}{|k\subrm{p}|} |k\subrm{p} - k\subrm{d} + k\subrm{c} | \ll \Gamma_{\text{Res}}^{{\text{4L}}}.
\end{equation}
Hence, the EIT linewidth $\Gamma_{\text{EIT}}^{\text{5L}}$ for the 3C5L-RAQR is no longer limited by Doppler mismatch $\Gamma_{\text{Res}}^{{\text{5L}}}$ but rather by the much smaller homogeneous broadening terms.
\begin{figure}[t]
	\centerline{\includegraphics[width=2.8in]{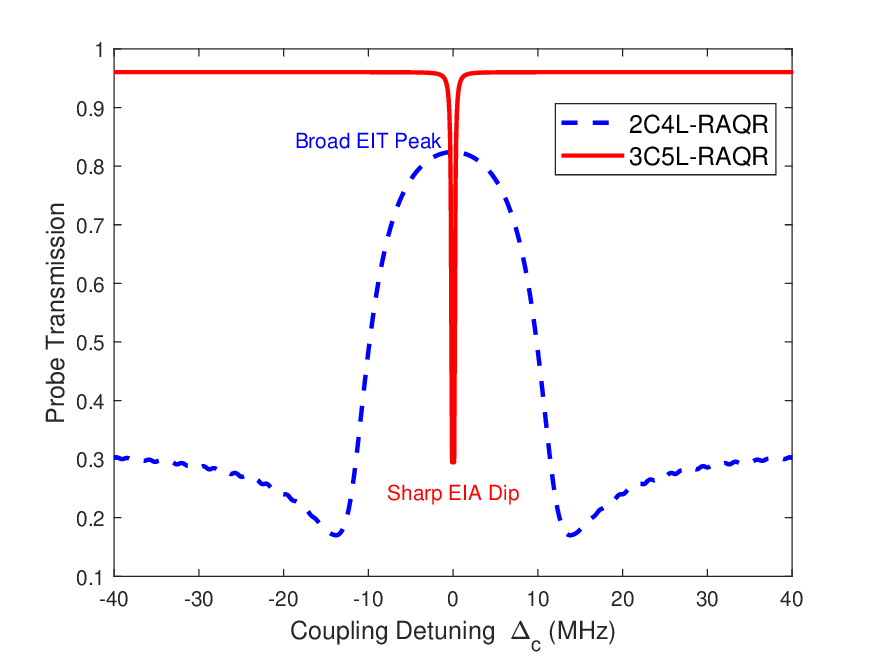}}
\caption{Transmission spectrum vs. coupling detuning $\Delta\subrm{c}$ at a vapor-cell temperature of $T_\text{atom}=290$ K.}
\label{fig:5L_3}
\end{figure}

 {In Fig. \ref{fig:5L_3}, we present a transmission spectrum of the steady-state probe transmission power, which is calculated by Eq. \eqref{eq:probe_power} composed of atomic susceptibility $\chi$ in Section II-C.} The 2C4L-RAQR exhibits a broad EIT window due to the uncancelled Doppler effects. Conversely, the 3C5L-RAQR maintains a sharp electromagnetically induced absorption (EIA) dip\footnote{In practice, {the presentation of EIT or EIA window in the 3C5L-RAQR depends on the relative intensity between the probe laser and the coupling laser. As presented in \cite[Fig. 2]{ref_prajapati_sensitivity}, the three-photon EIA is caused by using a weak probe field, while the EIT is caused by using a strong probe field. In this Section, we utilize a setup of weak probe field for deriving Eq. \eqref{eq:rho21_5level_continued_fraction}, thereby forming the three-photon EIA window.}}. The narrower EIA linewidth of the 3C5L-RAQR implies a significantly larger gradient with respect to frequency changes, which directly translates to a higher optical response for a given RF field strength. Note that the EIT/EIA linewidth is an important, but not exclusive factor governing the receiver sensitivity \cite{ref_prajapati_sensitivity}, while the actual sensitivity is jointly determined by the atomic transconductance, transmitted probe power, and the associated noise floor.

For the flexibility of spectrum access, while the 3C5L-RAQR uses a $P \to D$ transition for fairness with 2C4L-RAQR in Fig. \ref{fig:level_schemes}, the $S \to P \to S \to P$ excitation path can also be extended to the $S \to P \to D \to F$ scheme \cite{brown2023very}.
In this case, the 3C5L-RAQR populates F-states, which unlocks access to high-$l$ transitions, e.g., $\ket{nF} \to \ket{nG}$, $l=3 \to l=4$, preserving the advantage of UHF/VHF access. 


{\emph{\textbf{Remark 1:} Transitioning from the conventional 2C4L-RAQR to the 3C5L-RAQR architecture intuitively implies increased hardware complexity, as it requires three independent lasers to be phase-locked, frequency-stabilized, and spatially aligned. However, operating entirely within the mature red/infrared diode laser bands fundamentally bypasses the severe engineering bottlenecks of the 2C4L-RAQR setup. For instance, the 3C5L-RAQR avoids the highly unstable and power-hungry second-harmonic generation modules required for {blue/green} lasers \cite{zuo2025high}, rendering frequency stabilization and power-cost trade-offs highly practical with off-the-shelf components. \textbf{Consequently, the intrinsic physical compensations of the 3C5L-RAQR architecture dwarf the hypothetical complexities of an additional laser, making it practically viable for compact deployment and commercialization \cite{ref_infleqtion_sqywire}.}}}

\subsection{Baseband Signal Model of {3C5L-RAQR}}
\label{sec:signal_model}

In this section, we incorporate the specific 3C5L excitation modeling to formulate the tailored equivalent baseband signal model for the 3C5L-RAQR, where the superheterodyne method is employed to detect the weak RF signal-induced atomic response \cite{ref_gong_overview, ref_jing_superheterodyne}. This approach alleviates the detector noise inherent in traditional RAQRs that rely on the phase-insensitive Autler-Townes splitting (ATS) effect \cite{Holloway2014Broadband, ref_cui_mimo, ref_liu_continuous_freq}.

\subsubsection{Superheterodyne-Based Atomic Sensing}
In a typical superheterodyne RAQR architecture, a weak incoming RF signal field $U_x$ which carries the information at a carrier frequency $f_c$ is mixed with a strong coherent LO field $U_y$ at a nearby frequency $f_l$. This mixing process down-converts the signal to a low intermediate frequency (IF) $f_\delta = f_c - f_l$. 
The combined field $\Omega\subrm{RF}(t)$ can be decomposed into a strong static field $\Omega\subrm{LO}$ and a weak signal field $\Omega\subrm{sig}(t)$ that oscillates at the IF, which is given by 
\begin{align}
\Omega\subrm{RF}(t) &= \Omega\subrm{LO} + \Omega\subrm{sig}(t)=\Omega\subrm{LO} + \Omega_x \cos(2\pi f_\delta t + \theta_\delta),
\label{eq:rf_superposition}
\end{align}
where $\Omega_x$ is the Rabi frequency of the signal related to its electric field (E-field) $U_x$ by $\Omega_x = \mu_{45} U_x / \hbar \ll |\Omega\subrm{LO}|$ for the 5L $\ket{4}\to\ket{5}$ transition, and the IF phase is $\theta_\delta = \theta_x - \theta_y$, i.e., the signal phase $\theta_x$ relative to LO phase $\theta_y$. 

Suppose the input probe $P_0(t)$ is expressed as
\begin{align}
P_0(t) &= \sqrt{2\mathcal{P}_0} \cos(2\pi f\subrm{p} t + \phi_0), \label{eq:input_probe}
\end{align}
where $\mathcal{P}_0$, $f\subrm{p}$, and $\phi_0$ are initial power, frequency, and phase of input probe, respectively.

The probe beam $P_0(t)$ propagates through the vapor cell with length $d$ and is modulated by the susceptibility $\chi$. 
Given the wavelength of the probe laser $\lambda\subrm{p}$ and the length of the vapor cell $d$, the output probe beam $P(\Omega\subrm{RF}, t)$ in RAQRs is given by \cite{ref_gong_overview, holloway2017electric}
\begin{equation}
P(\Omega\subrm{RF}, t) = \sqrt{2\mathcal{P}_1(\Omega\subrm{RF})} \cos(2\pi f\subrm{p} t + \phi\subrm{p}(\Omega\subrm{RF})),
\label{eq:output_probe}
\end{equation}
where the output power $\mathcal{P}_1$ and phase $\phi\subrm{p}$ are given by
\begin{align}
{\mathcal{P}_1(\Omega\subrm{RF})} &= \mathcal{P}_0 e^{-\frac{2\pi d}{\lambda\subrm{p}} \Im(\chi(\Omega\subrm{RF})) }\label{eq:probe_power}, \\
\phi\subrm{p}(\Omega\subrm{RF}) &= \phi_0 + \frac{\pi d}{\lambda\subrm{p}} \Re(\chi(\Omega\subrm{RF})) \label{eq:probe_phase}.
\end{align}

Since $\Omega\subrm{RF}(t)$ is time-varying, both $\mathcal{P}_1$ and $\phi\subrm{p}$ are modulated in time. We linearize them around the strong LO bias $\Omega\subrm{LO}$ using a Taylor expansion, which can be expressed as
\begin{equation}
\chi(\Omega\subrm{RF}) \approx {\chi(\Omega\subrm{LO})} + \Omega\subrm{sig}(t) \chi'\subrm{s},
\label{eq:taylor_expansion}
\end{equation}
where $\chi'\subrm{s} = \left. \frac{\partial \chi}{\partial \Omega\subrm{RF}} \right|_{\Omega\subrm{LO}}$ is the atomic transconductance. 

{By intentionally driving the atomic ensemble with a strong LO field, the instantaneous variation induced by the incident signal remains a minuscule perturbation, allowing the superheterodyne RAQR to operate within a highly linear transconductance regime. 
Note that this linear approximation is strictly bounded by the RF nonlinearity of the atomic ensemble. When the incident RF signal becomes excessively strong, the atomic susceptibility $\chi(\Omega\subrm{RF})$ enters a nonlinear response regime, compressing the effective receiver gain.}

Then, substituting Eq. \eqref{eq:taylor_expansion} into Eq. \eqref{eq:probe_phase}, we obtain the phase modulation as
\begin{align}
\phi\subrm{p}(\Omega\subrm{RF}) &\approx \phi_0 + \frac{\pi d}{\lambda_p} \Re(\chi_0 + \Omega\subrm{sig}(t) \chi'\subrm{s}) \nonumber \\
&= {\left( \phi_0 + \frac{\pi d}{\lambda\subrm{p}} \Re(\chi_0) \right)} + {\frac{\pi d}{\lambda\subrm{p}} \Re(\Omega\subrm{sig}(t) \chi'\subrm{s})}.
\end{align}
Substituting it into Eq. \eqref{eq:probe_power} and using $e^x \approx 1+x$ for the small signal term, we obtain the amplitude modulation as
\begin{align}
\mathcal{P}_1(\Omega\subrm{RF}) &\approx \mathcal{P}_0 e^{-\frac{2\pi d}{\lambda\subrm{p}} \Im(\chi_0)} e^{-\frac{2\pi d}{\lambda\subrm{p}} \Im(\Omega\subrm{sig}(t) \chi'\subrm{s})} \nonumber \\
&\approx {\mathcal{P}_1(\Omega\subrm{LO})} \left( 1 - {\frac{2\pi d}{\lambda\subrm{p}} \Im(\Omega\subrm{sig}(t) \chi'\subrm{s})} \right).
\end{align}
Hence, the weak RF signal $\Omega\subrm{sig}(t)$ has successfully modulated both the amplitude and phase of the optical probe beam.

\subsubsection{BCOD-Based Photoelectric Detection}
To detect the strength of the optical probe beam modulated by the impinging RF field, we adopt the balanced coherent optical detection (BCOD) scheme \cite{ref_gong_overview}. BCOD utilizes a strong local optical beam, which helps to suppress the thermal noise generated by electronic components. Let the optical LO in BCOD be $P_l(t) = \sqrt{2\mathcal{P}_l} \cos(2\pi f\subrm{p} t + \phi_l)$, where $\mathcal{P}_l$ and $\phi_l$ are the power and phase of optical LO, respectively. BCOD mixes the modulated probe beam $P(\Omega\subrm{RF}, t)$ with the optical LO $P_l(t)$. The two beams are combined on a beam splitter, and the two outputs are sent to two photodiodes. The balanced detector subtracts the resulting photocurrents, which cancels out the large DC components and isolates the product of the two optical fields, forming the baseband photocurrent as \cite{ref_gong_overview}
\begin{align}
I\subrm{B}(t) &= 2 \alpha \sqrt{\mathcal{P}_l \mathcal{P}_1(\Omega\subrm{RF})} \cos(\phi_l - \phi\subrm{p}(\Omega\subrm{RF})) \label{eq:photocurrent},
\end{align}
where $\alpha$ is the photodetector responsivity.

The photocurrent signal $I\subrm{B}(t)$ is fed to a transimpedance amplifier (TIA) with load $R$ and a low noise amplifier (LNA) with power gain $G_{\text{PD}}$, and hence the output voltage $V\subrm{B}(t) = \sqrt{G_{\text{PD}}} R I\subrm{B}(t)$.
Substituting the photocurrent from Eq. \eqref{eq:photocurrent} and assuming a unit value $R$ of impedance load, we get
\begin{equation}
V\subrm{B}(t) = \sqrt{G_{\text{PD}}} \left[ 2 \alpha \sqrt{\mathcal{P}_l \mathcal{P}_1(\Omega\subrm{RF})} \cos(\phi_l - \phi\subrm{p}(\Omega\subrm{RF})) \right].
\end{equation}

We define $P\subrm{mix}(\Omega\subrm{RF}) \triangleq \sqrt{\mathcal{P}_1(\Omega\subrm{RF})} \cos(\phi_l - \phi\subrm{p}(\Omega\subrm{RF}))$ and linearize this measured term around $\Omega\subrm{LO}$ as
\begin{equation}
P\subrm{mix}(t) \approx P\subrm{mix}(\Omega\subrm{LO}) + \Omega\subrm{sig}(t) \left. \frac{d P\subrm{mix}}{d \Omega\subrm{RF}} \right|_{\Omega\subrm{LO}}.
\end{equation}

The total voltage can be rewritten as $V\subrm{B}(t) = V\subrm{DC} + \tilde{V}(t)$, where $V\subrm{DC} = 2 \alpha \sqrt{{G_{\text{PD}}}\mathcal{P}_l} P\subrm{mix}(\Omega\subrm{LO})$ is the large DC offset, and $\tilde{V}(t) = 2  \alpha \sqrt{{G_{\text{PD}}}\mathcal{P}_l} \Omega\subrm{sig}(t) P'\subrm{mix}(\Omega\subrm{LO})$ is the desired AC signal at the IF that contains the impinging microwave signal. 

\textbf{Lemma 2:} The desired AC component $\tilde{V}(t)$ at the IF $f_\delta$ in the 3C5L-RAQR is given by
\begin{equation}
\tilde{V}(t) = 2\alpha\sqrt{{G_{\text{PD}}}\mathcal{P}_l \mathcal{P}_1(\Omega\subrm{LO})} \kappa_1 \sin(\varphi_1){U_x} \cos(2\pi f_\delta t + \theta_\delta),
\label{eq:passband_voltage_simple}
\end{equation}
where the field-to-optical gain $\kappa_1$ and the total atomic-optical phase shift $\varphi_1$ are given by
\begin{equation}
\kappa_1 \triangleq \frac{\pi d \mu_{45}}{\lambda\subrm{p} \hbar} \left| \chi'\subrm{s} \right|,
\label{eq:kappa2_definition}
\end{equation}
\begin{equation}
\varphi_1= \phi_l - \phi\subrm{p}(\Omega\subrm{LO}) - \psi\subrm{p}
\label{eq:phi2_definition},
\end{equation}
where $\psi\subrm{p} = \text{arg}(\chi'\subrm{s})$.

\textit{Proof:} Please refer to Appendix \ref{app:derivation_BCOD} for more details. \hfill $\blacksquare$

\subsubsection{Equivalent Baseband Signal Model}
Next, we map the physical passband model in Eq. \eqref{eq:passband_voltage_simple} to the complex baseband communication model.
We define a complex baseband transfer function $\mathcal{H}\subrm{RAQR}$ to link the received baseband signal $x_b(t)$ to the output baseband voltage $\tilde{V}_b(t)$ via the linear relation, i.e., $\tilde{V}_b(t) = \mathcal{H}\subrm{RAQR} x_b(t)$. To find $\mathcal{H}\subrm{RAQR}$ in RAQRs, we relate the physical passband quantities to their complex baseband equivalents.
The received RF passband signal $x(t)$ with E-field amplitude $U_x$ and power $\mathcal{P}_x$ has a complex baseband representation $x_b(t) = \sqrt{\mathcal{P}_x} e^{j\theta_x}$. The E-field and power are related by $\mathcal{P}_x = U_x^2 A_e / (2 Z_0)$ with an equivalent aperture $A_e$, and $Z_0=1/(c\epsilon_0)$ is the impedance of free space. 
The physical output passband IF voltage $\tilde{V}(t)$ is related to its complex baseband equivalent $\tilde{V}_b(t)$ by
\begin{equation}
\tilde{V}(t) = \Re(\sqrt{2} \tilde{V}_b(t) e^{j2\pi f_\delta t}).
\label{eq:Vb_def}
\end{equation}
We define ${C_{\text{p}}}=2 \sqrt{G_{\text{PD}}} \alpha \sqrt{\mathcal{P}_l \mathcal{P}_1(\Omega_\text{LO})} \kappa_1 \sin(\varphi_1)$ and rewrite $\tilde{V}(t)$ in Eq. \eqref{eq:passband_voltage_simple} to match the form of Eq. \eqref{eq:Vb_def}, which can be expressed as
\begin{align}
\tilde{V}(t) &= C_{\text{p}} U_x \Re( e^{j(2\pi f_\delta t + \theta_\delta)} )= \Re (C_{\text{p}} U_x e^{j\theta_\delta} e^{j2\pi f_\delta t}).
\end{align}
By comparing this to Eq. \eqref{eq:Vb_def}, we can solve for $\tilde{V}_b(t)$:
\begin{equation}
\sqrt{2} \tilde{V}_b(t) = C_{\text{p}} U_x e^{j\theta_\delta} = C_{\text{p}} U_x  e^{j(\theta_x - \theta_y)}.
\end{equation}
Substituting $U_x = \sqrt{2 Z_0 / A_e} |x_b(t)|$ and $e^{j\theta_x} = x_b(t) / |x_b(t)|$, we have
\begin{align}
&\tilde{V}_b(t) = \frac{1}{\sqrt{2}} C_{\text{p}} \sqrt{\frac{2 Z_0}{A_e}} |x_b(t)|  \frac{x_b(t)}{|x_b(t)|} e^{-j\theta_y} \nonumber \\
& = \left[ \frac{2 \sqrt{G_{\text{PD}}} \alpha \sqrt{Z_0 \mathcal{P}_l \mathcal{P}_1(\Omega\subrm{LO})} \kappa_1 \sin(\varphi_1)}{\sqrt{A_e}} e^{-j\theta_y} \right] x_b(t).
\end{align}
The complete complex baseband transfer function of the RAQR is given by
\begin{equation}
\mathcal{H}\subrm{RAQR} = \frac{2\alpha  \sqrt{Z_0{G_{\text{PD}}} \mathcal{P}_l \mathcal{P}_1(\Omega\subrm{LO})} \kappa_1 \sin(\varphi_1)}{\sqrt{A_e}} e^{-j\theta_y}.
\label{eq:H_RAQR_derived}
\end{equation}

Let the transmitted baseband signal of the RF transmitter be $s_b$ and the wireless channel be $h$. The signal at the receiver aperture of the RAQR is $x_b = {\sqrt{A_e}} h s_b$. Based on the detected voltage $\tilde{V}_b$, the end-to-end baseband signal model of RAQR can be expressed as
\begin{align}
\tilde{V}_b &= {\sqrt{A_e}} {\mathcal{H}\subrm{RAQR}} {h} s_b + w \triangleq  {\sqrt{\beta}} \Phi {h} s_b + w,
\label{eq:siso_model}
\end{align}
where $\beta = 4 \alpha^2{G_{\text{PD}}} Z_0 \mathcal{P}_l \mathcal{P}_1(\Omega\subrm{LO}) \kappa_1^2 \sin^2(\varphi_1)$ denotes the effective power gain of the RAQR and $\Phi = \bar {\mathcal{H}}\subrm{RAQR} / |\bar {\mathcal{H}}\subrm{RAQR}|= e^{-j\theta_y}$ denotes the phase response by locking the optical LO phase $\phi_l$ to maximize the signal amplitude, i.e., setting $\sin(\varphi_1) \approx 1$. In the RAQR, $w \sim \mathcal{CN}(0, \sigma_w^2(s_b))$ denotes the signal-related noise\footnote{{While the underlying QPN and PSN processes are intrinsically signal-dependent and non-Gaussian, the superheterodyne detection regime allows for a linearized treatment of the noise variance. In this high-photon-flux limit, the accumulation of photon statistics and the dominance of the LO power justify the application of the central limit theorem \cite{ref_gong_overview}. Consequently, the conditional Gaussian model serves as an effective and physically consistent analytical approximation within the linear dynamic range of RAQRs.}}. The noise power $\sigma_w^2(s_b)= N\subrm{QPN} + N\subrm{PSN} + N\subrm{ITN}$ is the sum of quantum projection noise (QPN) $N\subrm{QPN}$, photon shot noise (PSN) $N\subrm{PSN}$, and intrinsic thermal noise (ITN) from electronics $N\subrm{ITN}$. 

In particular, the RAQR couples to the environmental electromagnetic vacuum, introducing the widely known black-body radiation (BBR) noise caused by the thermal field at environmental temperature $T_{\text{env}}$ on Rydberg atom sensors. Since RAQRs operate as coherent quantum sensors, the incoherent thermal fields do not drive the coherent Rabi oscillations required for signal detection \cite{kaur2025impact}. Instead, BBR induces random transitions between Rydberg states, leading to population redistribution and decoherence. The impact of BBR can be modeled as a thermal decoherence rate $\Gamma_{\text{BBR}}$ that broadens the Rydberg states, which is given by \cite{kaur2025impact, cooke1980effects}
\begin{equation}
    \Gamma_{\text{BBR}}\approx \frac{4 \alpha_{\text{eff}}^3 k_B T_{\text{env}}}{3 n_{\text{eff}}^2 \hbar},
    \label{eq:gamma_bbr}
\end{equation}
where $k_B$ is a Boltzmann constant. $\alpha_{\text{eff}}$ is a fine-structure constant and $n_{\text{eff}}$ is an effective principal quantum number.

\emph{\textbf{Remark 2:}} \emph{In classical RF receivers, the BBR noise is also termed thermal noise, where BBR manifests as the typical additive white Gaussian noise (AWGN), and hence the theoretical noise floor is fixed at $-174$ dBm/Hz due to thermal power density $k_B T_{\rm{env}}$ \cite{bussey2022quantum}. However, the impact of BBR in RAQRs is not as an additive noise source, but as a thermal decoherence rate that broadens the Rydberg states, where the authors of \cite{kaur2025impact} observed that the sensitivity
is only reduced by 0.1 \% at $T_\text{env}$ = 773.15 K based on practical experiments. \textbf{Indeed, treating BBR simply as an AWGN source would mathematically bound both atomic and electronic receivers to the same fundamental thermal noise floor, negating the quantum advantage of the RAQR.}}

    For the other dominated noise sources, the ITN power over the
receiver bandwidth $B_{\text{RAQR}}$ is given by
    \begin{equation}
        N_{\text{ITN}} = k_B T_{\text{PD}}  {G_{\text{PD}}}B_{\text{RAQR}},
    \end{equation}
    where $T_{\text{PD}}$ is the equivalent noise temperature and $B_{\text{RAQR}}$ denotes the communication bandwidth of RAQRs. 
In the BCOD scheme, the signal power scales with strong $\mathcal{P}_l$ while $N_{\text{ITN}}$ remains constant, resulting in the optical noise dominating thermal noise. 
    Then, the PSN power is given by
    \begin{align}
        N_{\text{PSN}} &= 2 q_0 {G_{\text{PD}}} \alpha (\mathcal{P}_l + \mathcal{P}_1)B_{\text{RAQR}},
        \label{PSN}
    \end{align}
    where $q_0$ is the elementary charge. 
    Moreover, the QPN power can be expressed as
    \begin{equation}
        N_{\text{QPN}} = \frac{\beta}{2Z_0} \left( \frac{\hbar \sqrt{\Gamma_{\text{EIT}}}}{\mu_{45} \sqrt{N_m}} \right)^2  B_{\text{RAQR}},
        \label{QPN}
    \end{equation}
    where $N_m= \pi r_0^2 d N_a$ denotes the number of Rydberg-state atoms within the probe-laser beam with beam radius $r_0$.

\emph{\textbf{Remark 3:}} \emph{\textbf{While the superheterodyne RAQRs predominantly emphasize the conventional optical mixing constraint $\Omega_{\rm{sig}} \ll \Omega_{\rm{LO}}$ to construct the linear response regime, the fundamental atomic saturation limit $\Omega_{\rm{sig}} \ll \Gamma_{\rm{EIT}}$  bounded by the EIT linewidth cannot be overlooked.} In practical high-power scenarios, violating these conditions leads to two distinct nonlinearity limitations: 1) the atomic saturation effect when $\Omega_{\rm{sig}} \approx \Gamma_{\rm{EIT}}$, where strong fields depopulate the ground state and compress the susceptibility; and 2) the LO-signal mixing saturation when $\Omega_{\rm{sig}} \approx \Omega_{\rm{LO}}$, where the strong signal breaks the linear heterodyne assumption. \textbf{Consequently, the dynamic range of RAQRs naturally dictates a dual-regime paradigm: the superheterodyne method is suitable for highly-sensitive weak-signal reception, whereas the ATS effect remains the mandatory mechanism for strong-field detection.}}

\section{Performance Comparison: 2C4L-RAQR vs. 3C5L-RAQR}
\label{sec:performance_analysis}

{Building upon the established foundational models in Section II, this section presents a quantitative performance analysis and comparison between the conventional 2C4L-RAQR and the exhibited 3C5L-RAQR.}

\subsection{ Scaling Laws of SNR}

Consider a Rayleigh fading channel denoted by the complex coefficient $h \sim \mathcal{CN}(0, 1)$ and leveraging the RAQR signal model derived in Eq. \eqref{eq:siso_model}, we can establish an end-to-end relationship between the received baseband signal power $P_S$ at the RAQR and the transmit power $P_t$ at the RF transmitter, which is given by \cite{Balanis2016Antenna}
\begin{align}
P_S = \beta |\Phi|^2 {\frac{P_t G_t}{4 \pi L^2}}|h|^2= P_t \frac{G_t \kappa_2|h|^2}{4 \pi L^2} |\chi'\subrm{s}|^2,
\label{eq:power_link_budget}
\end{align}
where $G_t$ is the transmit antenna gain and $L$ is the distance between the transmitter and the receiver. $\kappa_2 \triangleq 4 {G_\text{PD}}\alpha^2 Z_0 \mathcal{P}_l \mathcal{P}_1 \left( \frac{\pi d \mu_{45}}{\lambda_p \hbar} \right)^2 \sin^2(\varphi_1)$ aggregates the optoelectronic conversion gain and optical path parameters. 

\textbf{Lemma 3:} The analytical solution of atomic transconductance $ \chi'\subrm{s, \text{5L}}$ under weak-probe field approximation in the 3C5L-RAQR is given by
\begin{align}
 \chi'\subrm{s, \text{5L}}= \left( \frac{K \Omega_{\text{LO}}}{2} \right) \left[ \frac{ -(i\Omega\subrm{p}/2) (|\Omega\subrm{d}|^2/4) (|\Omega\subrm{c}|^2/4) }{ f_2^2 f_3^2 f_4^2 f_5 } \right],
 \label{transconductance}
\end{align}
where $K = \frac{2N_a |\mu_{12}|^2}{\epsilon_0 \hbar \Omega\subrm{p}}$ and the nested denominators are
\begin{align}
&f_5 = (i\Delta_5 - \Gamma_{51}), \\
&f_4 = (i\Delta_4 - \Gamma_{41}) + |\Omega_{\text{RF}}|^2 / (4 f_5), \\
&f_3 = (i\Delta_3 - \Gamma_{31}) + |\Omega\subrm{c}|^2 / (4 f_4), \\
&f_2 = (i\Delta_2 - \Gamma_{21}) + |\Omega\subrm{d}|^2 / (4 f_3). 
\end{align}

\textit{Proof:} Please refer to Appendix \ref{app:derivation_transconductance} for more details. \hfill $\blacksquare$

The atomic transconductance $\chi'\subrm{s, \text{5L}}$ is determined by the nested denominators $f_n,n \in \{2, \ldots ,5\}$. Considering a thermal atom with velocity ${v}$, $f_n$ can be rewritten as 
\begin{align}
f_n(v) = i(\Delta_n - \Delta_k {v}) - \Gamma_{n1}. 
\end{align}
where $\Delta_k$ denotes the wavevector mismatch of lasers.

In the conventional 2C4L-RAQR, the large wavelength mismatch results in a significant residual wavevector $\Delta k_\text{4L}$. The denominators $f_3(v)$ and $f_4(v)$ are dominated by the Doppler shift term $i(\Delta_k v)$ with $\Delta_k=|k_p - k_c|$. Following Eq. \eqref{eq:gamma_res_4L}, the effective EIT linewidth is dominated by $\Gamma^{\text{4L}}_{\text{Res}} = \frac{\Gamma_{2}}{|k_p|}\Delta_k$, which is much larger than the original decay rates $\Gamma_{n1}$. Consequently, the atomic sensitivity of the 2C4L-RAQR is limited by this MHz-scale broadening rather than the natural linewidths. For the convenience of performance analysis by exhibiting the upper bound of the atomic transconductance $\chi'\subrm{s}$, suppose the resonant frequency and the weak-field power broadening effects \cite{ref_doppler_mismatch, ref_prajapati_sensitivity}, i.e., $|\Omega|^2/|f_{n+1}| \ll \Gamma_{\text{n1}}$, we have
\begin{align}
\chi'\subrm{s, \text{4L}}  \propto \frac{1}{|f_2^2 f_3^2 f_4|} &\propto \frac{1}{{\Gamma_{21}^2} \left( {\Gamma_{31}} + \Gamma_{\text{Res}}^{\text{4L}} \right)^2 \left( {\Gamma_{41}} + \Gamma_{\text{Res}}^{\text{4L}} \right)},
\label{eq:scaling_4L_rigorous}
\end{align}
where the term $\Gamma_{\text{Res}}^{\text{4L}}$ dominates the denominator, significantly suppressing the atomic response. 

The 3C5L-RAQR architecture utilizes a three-photon resonance condition, i.e., $\vec{k}_p + \vec{k}_d + \vec{k}_c \approx 0$, to cancel the Doppler effect, where the residual Doppler linewidth $\Gamma^{\text{5L}}_{\text{Res}}= \frac{\Gamma_{2}}{|k_p|} |k_p - k_d + k_c |$ is much smaller than the $\Gamma^{\text{4L}}_{\text{Res}}$. The terms $f_n$ are dominated by the natural linewidths $\Gamma_{n1}$. The atomic sensitivity scales as
\begin{align}
\chi'_{s, \text{5L}} \propto \frac{1}{|f_2^2 f_3^2 f_4^2 f_5|} \propto \frac{1}{\Gamma_{21}^2 \Gamma_{31}^2 (\Gamma_{41}+\Gamma^{\text{5L}}_{\text{Res}})^2 (\Gamma_{51}+\Gamma^{\text{5L}}_{\text{Res}})}.
\end{align}

\emph{\textbf{Remark 4:}} \emph{Compared to 2C4L-RAQR, the 3C5L-RAQR introduces an additional linewidth term $\Gamma_{31}$ in the denominator, originating from the extra intermediate transition excited by the dressing laser. The sensitivity scaling is determined by the product of the linewidths along the excitation path. However, the bottleneck in 2C4L-RAQR is the squared MHz-scale residual Doppler width $\Gamma^{\rm{4L}}_{\rm{Res}}$. \textbf{Although the 3C5L-RAQR adds an extra term $\Gamma_{31}$, it  replaces the MHz-scale term $\Gamma^{\rm{4L}}_{\rm{Res}}$ with a squared kHz-scale term $\Gamma^{\rm{5L}}_{\rm{Res}}$, translating into a significant enhancement in macroscopic sensitivity.}}

In the RAQR, the sensitivity of RF field detection is inversely proportional to the EIT linewidth ${\Gamma_{\text{EIT}}}$, and hence the SNR evaluation of the RAQR is also related to ${\Gamma_{\text{EIT}}}$. The macroscopic EIT linewidth ${\Gamma_{\text{EIT}}}$ that accounts for natural decay rate \cite{ref_doppler_mismatch}, BBR-induced broadening \cite{kaur2025impact}, and other external broadening mechanisms \cite{meyer2018digital, holloway2017electric, vsibalic2017arc}, can be formulated as
\begin{equation}
    \Gamma_{\text{EIT}} = \Gamma_{\text{Res}} + \frac{ \Gamma_\text{Ryd}^{\text{nat}} + \Gamma_{\text{BBR}}}{2} +  \frac{v_\text{th}}{r_0} +  {{\Omega_\text{AT}}} +\Gamma_{\text{d}},
    \label{eq:gamma_eit_final}
\end{equation}
where $\Gamma_\text{Ryd}^{\text{nat}}$ denotes the natural decay rate of the sensing Rydberg state, i.e., $\Gamma_\text{Ryd}^{\text{nat}}= \Gamma_3^{\text{4L}}$ for 2C4L-RAQR and $\Gamma_\text{Ryd}^{\text{nat}}= \Gamma_4^{\text{5L}}$ for 3C5L-RAQR. { 
The term $\Omega_\text{AT}$ describes the power-broadening contribution that generally depends on all driving
fields \cite{meyer2018digital, holloway2017electric}, i.e., $\Omega_p$, $\Omega_c$, $\Omega_{\mathrm{LO}}$, and $\Omega_d$ for 3C5L-RAQR.
The term $\frac{v_\text{th}}{r_0}$ denotes transit-time broadening of thermal atoms, where $v_{\text{th}}= \sqrt{{2 k_B T_\text{atom}}/{m}}$ denotes the most probable transverse velocity component for the atom with a mass of $m$ and is orthogonal to $v_z$ in Eq. \eqref{ff}. The dephasing term $\Gamma_{\text{d}}$ includes other external broadening effects, e.g., atom collision and laser linewidths \cite{vsibalic2017arc}. For the residual Doppler broadening, the term $\Gamma_{\text{Res}}^{\text{4L}}$ is significant due to wavevector mismatch for 2C4L-RAQR, whereas in 3C5L-RAQR, $\Gamma_{\text{Res}}^{\text{5L}}$ is small owing to the Doppler-cancellation configuration. 
As a result, the EIT linewidth of the 2C4L-RAQR is dominated by $\Gamma_{\text{EIT}}^{\text{4L}} \approx \Gamma_{\text{Res}}^{\text{4L}}\gg \Gamma_{\text{EIT}}^{\text{5L}}$ \cite{ref_prajapati_sensitivity}. }

The SNR $\gamma^j_{\text{RAQR}}, j \in \{\text{4L, 5L}\}$ for the RAQR is given by
\begin{equation}
\gamma^j_{\text{RAQR}} = \frac{P_S}{N_{\text{PSN}} + N_{\text{QPN}} + N_{\text{ITN}}}.
\end{equation}
\textbf{Key Takeaway 1}: The SNR scaling law in the RAQR leads to two operating regimes:
\begin{itemize}
    \item \textbf{PSN-Limited Regime:} 
    If $N_{\text{PSN}}$ typically dominates at high LO power in Eq. \eqref{PSN}, the SNR advantage of the 3C5L-RAQR comes from signal enhancement:
    \begin{align}
\frac{\gamma_{\text{RAQR}}^{\text{5L}}}{\gamma_{\text{RAQR}}^{\text{4L}}} \approx \frac{P_S^{\text{5L}}/ N^{\text{5L}}_{\text{PSN}}}{P_S^{\text{4L}}/ N^{\text{4L}}_{\text{PSN}}} \propto { \frac{\mathcal{P}_\text{0}^{\text{5L}}}{\mathcal{P}_\text{0}^{\text{4L}}} } \left| \frac{\chi'_{s, \text{5L}} }{\chi'_{s, \text{4L}} } \right|^2 \propto { \frac{\mathcal{P}_\text{0}^{\text{5L}}}{\mathcal{P}_\text{0}^{\text{4L}}} } \left( \frac{\Gamma_{\text{EIT}}^{\text{4L}}}{\Gamma_{\text{EIT}}^{\text{5L}}} \right)^4.
       \label{x}
    \end{align}

    \item \textbf{QPN-Limited Regime:}
    If the system is optimized to reach the atomic limit dominated by $N_{\text{QPN}}$, the SNR gains between 5L and 4L RAQRs become
    \begin{align}
         \frac{\gamma_{\text{RAQR}}^{\text{5L}}}{\gamma_{\text{RAQR}}^{\text{4L}}} \approx \frac{P_S^{\text{5L}} / N^{\text{5L}}_{\text{QPN}} }{P_S^{\text{4L}} / N^{\text{4L}}_{\text{QPN}}}\propto  \frac{\beta^{\text{5L}} / (\beta^{\text{5L}} \Gamma_{\text{EIT}}^{\text{5L}})}{\beta^{\text{4L}} / (\beta^{\text{4L}} \Gamma_{\text{EIT}}^{\text{4L}})} \propto \frac{\Gamma_{\text{EIT}}^{\text{4L}}}{\Gamma_{\text{EIT}}^{\text{5L}}}.
         \label {xx}
    \end{align}
    
\end{itemize}

\begin{figure}[t]
	\centerline{\includegraphics[width=2.8in]{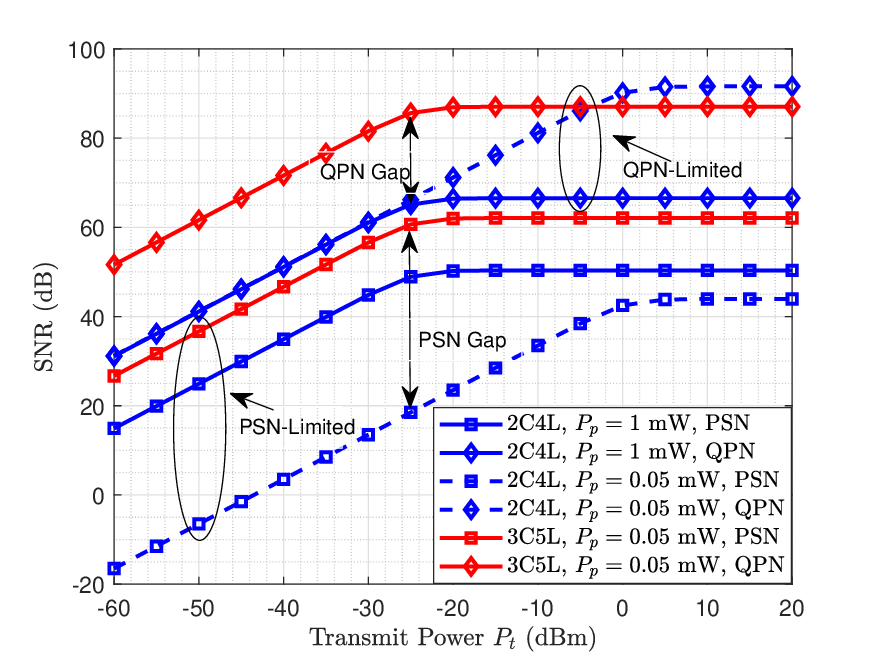}}
\caption{SNR performance under different noise regimes of RAQRs.}
\label{noise}

\end{figure}

Fig. \ref{noise} decomposes the received SNR into two fundamental PSN- and QPN-limited regimes, where the 2C4L- and 3C5L-RAQRs are evaluated under different probe laser powers $\mathcal{P}_0^{\text{4L}}$. When operating under an identical probe laser power $\mathcal{P}_0^{\text{4L}} = 0.05$ mW, the 3C5L-RAQR architecture demonstrates a sensitivity advantage, while the 2C4L-RAQR suffers from an obvious collapse in the PSN-limited regime. This enormous PSN gap visually corroborates the fourth-power scaling penalty derived in Eq. \eqref{x}. Due to the severe residual Doppler mismatch in the 2C4L configuration, the macroscopic transconductance is strongly suppressed, leaving the weak signal entirely submerged by PSN. {We observe that under the QPN-limited regime, the 3C5L-RAQR response enters gain compression earlier. This is because the suppression of residual Doppler broadening improves the small-signal SNR, while simultaneously reducing the RF field range over which the LO-biased atomic transduction remains linear.}
To mitigate this severe signal loss, a conventional workaround for the 2C4L-RAQR is to drastically increase the optical driving power, e.g., $\mathcal{P}_0^{\text{4L}} = 1$ mW.  While this brute-force power injection amplifies the transconductance, it induces severe power broadening to increase $\Gamma^\text{4L}_{\text{EIT}}$ and breaks the weak-probe approximation. 


\subsection{Communication Bandwidth and {Achievable Rate}}
    
For the RAQR, the communication bandwidth $B_{\text{RAQR}}$ is bounded by the transient response speed of the system, i.e., the transient relaxation time $\tau_f$ required for the Rydberg atoms to establish a new steady state upon the signal variations \cite{meyer2018digital}. In practical room-temperature vapor cells, this transient response is governed not only by the internal cascade decay of atomic states but also macroscopically constrained by the transit time of atoms flying across the laser beam profile and the power broadening of lasers or other signal fields. 
In Section IV-C, we will elaborate the exact numerical solution of $B_{\text{RAQR}}^j$ from the time-dependent Lindblad master equation. 
The {achievable rate} of RAQRs is expressed as\footnote{In fact, the signal-dependent noise in RAQRs is fundamentally coupled with the RF signal state, i.e., the $ \mathcal{P}_1$-related PSN in Eq. \eqref{PSN}, $\Gamma_\text{EIT}$-related QPN in Eq. \eqref{QPN} as well as non-AWGN BBR noise. Hence, the classical Shannon channel capacity formula that requires standard AWGN distribution is no longer applicable for RAQRs \cite{8450017}. {Considering the weak $\Omega_{\text{sig}}$ and the strong $\Omega_{\text{LO}}$, this work follows the Gaussian achievable rate as a tractable performance metric and incorporates the effective SNR $\gamma^j_{\text{RAQR}}$ term.}}
\begin{align}
C_{\text{RAQR}}^j = B_{\text{RAQR}}^{j} \mathbb{E}_{{h}} \left[ \log_2  \left( 1+ {\gamma_{\text{RAQR}}^j} \right) \right].
\end{align}

To evaluate the relative performance of different receiver architectures, we consider a classical RF receiver utilizing a standard conductor antenna. The isotropic antenna captures power through its effective aperture $A_{\text{CL}}= G_{{r}} \frac{\lambda_{\text{RF}}^2}{4\pi}$, where $G_{{r}}$ denotes the antenna gain. Given $\mathbb{E}[|h|^2]=1$, the averaged received signal power $P_{S, \text{CL}}$ after the LNA is given by
    \begin{align}
        P_{S, \text{CL}} &=  \frac{P_t G_t}{4 \pi L^2}  G_{\text{LNA}} A_{\text{CL}}= P_t {\frac{G_t G_{r} G_{\text{LNA}} \lambda_{\text{RF}}^2}{(4 \pi L)^2}},
    \end{align}
where $G_{\text{LNA}}$ denotes the gain of the LNA.

In this work, we refer to the noise power modeling in \cite{ref_gong_overview} for a classical RF receiver, which is given by
\begin{equation}
    P_{N, \text{CL}} = k_B T_{\text{env}} F G_{\text{LNA}} B_{\text{CL}}.
    \label{eq:noise_classical_explicit}
\end{equation}
where $F$ is the noise factor of the holistic system, and $B_{\text{CL}}$ is the bandwidth of the classical RF receiver.

    The SNR for the classical RF receiver is expressed as
    \begin{equation}
        \gamma_{\text{CL}} = \frac{P_{S, \text{CL}}}{P_{N, \text{CL}}} = \frac{G_t G_{{r}} \lambda_{\text{RF}}^2 P_t}{(4 \pi L)^2 k_B T_{\text{env}} FB_{\text{CL}}}.
        \label{eq:SNR_CL_Exact}
    \end{equation}

The achievable rate of the classical RF receiver is given by
    \begin{align}
C_{\text{CL}} &= B_{\text{CL}} \mathbb{E}_{{h}} \left[ \log_2  \left(1+ {\gamma_{\text{CL}}}\right) \right]. 
    \end{align}

Note that for the classical RF receiver, the bandwidth $B_{\text{CL}}$ is a designable parameter to determine the filter width, whereas for the RAQR, the bandwidth $B_{\text{RAQR}}^j$ is physically constrained by the atomic parameters, laser intensity and RF signal field, which is generally far smaller than that of a classical RF receiver. 

\subsection{Spectrum Access Range}
\label{ssec:freq_range}
{In classical RF engineering, maintaining high sensitivity at lower frequencies requires massive macroscopic antennas. Miniaturizing classical receivers in these bands typically results in severe sensitivity degradation due to a rapidly diminishing effective aperture. However, in the quantum regime of RAQRs, the fundamental scaling laws dictate a different paradigm.} Specifically, in RAQR, the choice of excitation pathway dictates the accessible RF frequencies.
The energy of a Rydberg state is $E_{n,l} = - R_y / (n - \delta_l)^2$, where $R_y$ is a Rydberg constant and $\delta_l$ is the quantum defect that decreases rapidly with $l$ \cite{vsibalic2017arc, steck2025rubidium}.
In 2C4L-RAQR, the atomic excitation path is $S \to P \to D$. For a transition $\ket{nD} \to \ket{(n+1)P}$ in RAQRs, the detectable frequency of the incident RF signal is given by\cite{zhang2023quantum}:
    \begin{equation}
    f_c \approx \frac{R_y}{2\pi \hbar} \left| \frac{1}{(n - \delta_D)^2} - \frac{1}{(n+1 - \delta_P)^2} \right|.
    \label{fre}
    \end{equation}
    The large difference between $\delta_P$ and $\delta_D$ results in $f_c$ in the high-frequency range. The large energy difference restricts 2C4L-RAQR to high-GHz or THz detection. 

\textbf{Key Takeaway 2}: The exhibited 3C5L-RAQR provides a flexible platform for accessing different frequency bands.
    \begin{itemize}
    \item \textbf{Microwave Access:} 
    By following an atomic transition pathway similar to that of the 2C4L-RAQR,  i.e., $S \rightarrow P \rightarrow S \rightarrow P$ in Section II-B, the 3C5L-RAQR can access microwave frequencies in the GHz/THz range.
    
    \item \textbf{UHF/VHF Access:} The alternative schemes, e.g., $S \rightarrow P \rightarrow D \rightarrow F$, can be designed to populate high orbital angular momentum F-states. This unlocks access to high-quantum transitions, e.g., $\ket{nF} \to \ket{nG}$ \cite{brown2023very}, where the signal frequency is given by
    \begin{equation}
    f_c \approx \frac{R_y}{2\pi \hbar} \left| \frac{1}{(n - \delta_F)^2} - \frac{1}{(n - \delta_G)^2} \right|.
    \label{eq:exact_freq}
    \end{equation}
    
    Since $\delta_F$ and $\delta_G$ are tiny \cite{vsibalic2017arc}, the energy difference is minimal. Using $1/(n-a)^2 - 1/(n-b)^2 \approx 2(a-b)/n^3$:
    \begin{equation}
    f_c \approx \frac{ R_y |\delta_F - \delta_G|}{\pi \hbar n^3}.
    \end{equation}
    We can observe that for transitions between high-angular-momentum states with similar quantum defects, the resonant frequency scales inversely with the principal quantum number $n^{-3}$ and is proportional to the differential quantum defect $|\delta_F - \delta_G|$  \cite{brown2023very}. In Fig. \ref{fig:access} of Section V, we show that the exhibited 3C5L-RAQR is a unified platform for both microwave and UHF/VHF bands.
    \end{itemize}



\section{Exact Numerical Modeling of Open Quantum Dynamics for Practical RAQRs}
\label{sec:exact_modeling}

\begin{figure}[t]
	\centerline{\includegraphics[width=2.8in]{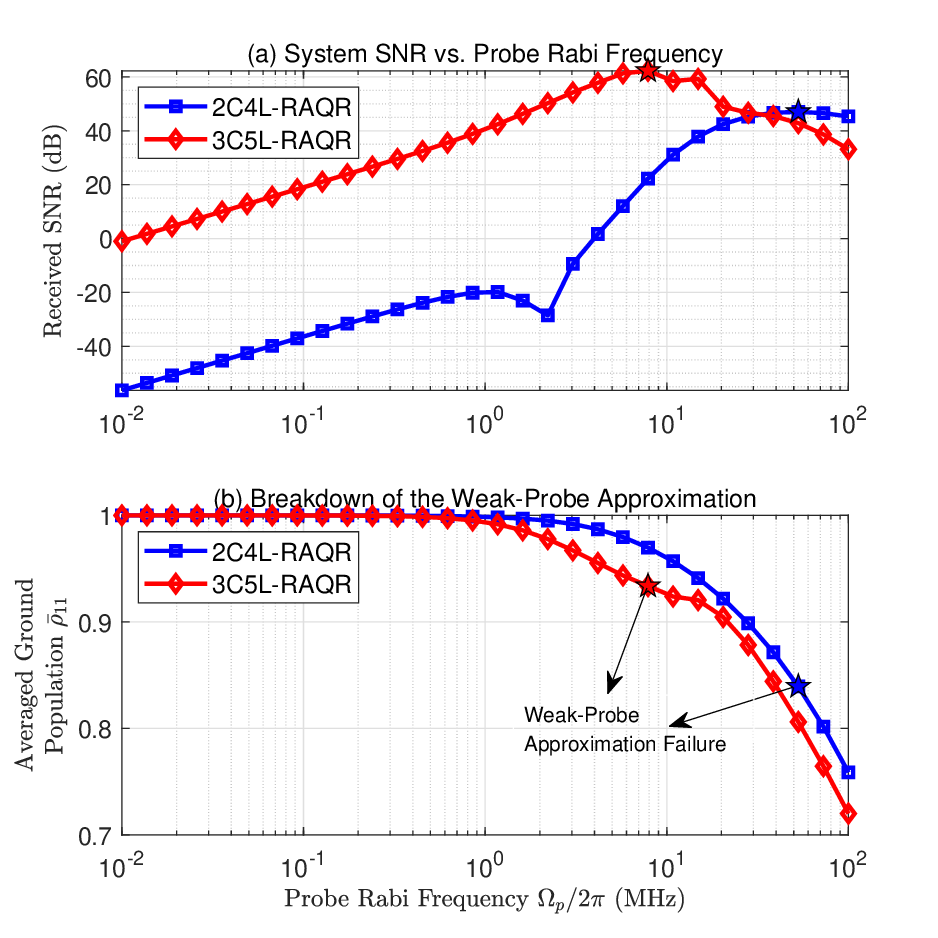}}
\caption{Probe Rabi frequency vs. SNR and ground-state atom population.}
\label{Probe}
\end{figure}

As established in Section II and III, the analytical model of steady-state atomic coherence $\rho_{21}$ relies on the weak-probe approximation $\bar{\rho}_{11} \approx 1$. However, the conventional weak-probe approximation is fundamentally sub-optimal for maximizing sensitivity of RAQRs \cite{meyer2021optimal}. As shown in Fig. \ref{Probe}, the $\bar{\rho}_{11} \approx 1$ approximation breaks down near the SNR peak, where the increase of probe laser intensity will cause the optical pumping of ground-state atoms. To rigorously evaluate the communication performance of RAQRs, in this section, we  transition to a full computational physics framework, providing the exact numerical solution for open quantum systems. 

\subsection{Transit-Time Relaxation and Probability Conservation}
In the exact numerical modeling, we recall the Lindblad master equation in Eq. \eqref{eq:rx_master_eq} as
\begin{equation}
\frac{d\hat{\rho}}{dt} = -\frac{i}{\hbar}[\hat{H}, \hat{\rho}] + \mathcal{L}_{\text{decay}}(\hat{\rho}) + \mathcal{L}_{\text{transit}}(\hat{\rho}),
\label{eq:master_eq_full}
\end{equation}
where $\mathcal{L}_{\text{decay}}$ denotes the intrinsic spontaneous emission and collision-induced dephasing. $\mathcal{L}_{\text{transit}}(\hat{\rho})$ denotes the transit-time relaxation, which describes thermal atoms traversing and exiting the finite spatial cross-section of the driving laser beams \cite{mohapatra2007coherent,vsibalic2017arc}. 
{To account for practical experimental imperfections rather than assuming ideal zero-linewidth lasers, the off-diagonal elements of the dissipation superoperator $\mathcal{L}_\text{decay}(\hat{\rho})$ incorporate a phenomenological dephasing matrix $\Gamma_\text{coh} = \frac{\Gamma_\text{pop}^{(i)} + \Gamma_\text{pop}^{(j)}}{2} + \Gamma_{\text{d}}, (i \neq j)$, where $\Gamma_\text{pop}^{(i)}$ denotes the intrinsic spontaneous emission and BBR-induced population decay rate of state $|i\rangle$. The residual pure dephasing rate $\Gamma_{\text{d}}$ is introduced to account for the finite laser linewidths, phase noise and residual atom-atom collisional broadening, which is physically identical to the macroscopic dephasing term $\Gamma_{\text{d}}$ in the analytical EIT linewidth model of Eq. \eqref{eq:gamma_eit_final}. }

To obtain the exact steady-state coherence under $\frac{d\hat{\rho}}{dt} = 0$ without relying on the weak-probe approximation, we employ the Liouvillian superoperator formalism \cite{steck2007quantum}. 
By mapping the $N \times N$ density matrix $\hat{\rho}$ onto a column vector $| \rho \rangle\rangle$ of dimension $N^2 \times 1$  where $N=5$ for the 3C5L-RAQR, the matrix differential equation is transformed into a linear algebraic system \cite[Sec 4.1.2]{steck2007quantum}:
\begin{equation}
\mathbf{L} | \rho \rangle\rangle = 0,
\end{equation}
where $\mathbf{L}$ is the $25 \times 25$ non-Hermitian Liouvillian matrix. 

For a room-temperature vapor cell, the finite transit time of atoms crossing the localized laser beams constitutes a primary source of decoherence, where atoms exit the interaction region at a characteristic rate $\gamma_t = v_{\text{th}} / r_0$ \cite{sagle1996measurement,hu2023improvement}. 
Critically, to maintain a constant atomic density within the sensing volume, the lost probability amplitude from exiting atoms must be precisely replenished by fresh and unpolarized atoms entering the beam in the ground state $|1\rangle$. We model the open-system boundary condition as \cite{meyer2021optimal}
\begin{equation}
\mathcal{L}_{\text{transit}}(\hat{\rho}) = -\gamma_t \hat{\rho} + \gamma_t \text{Tr}(\hat{\rho})  |1\rangle\langle 1|.
\label{eq:transit_model}
\end{equation}

Since the system is trace-preserving, i.e., $\text{Tr}(\hat{\rho}) = 1$, the Liouvillian matrix $\mathbf{L}$ is singular, possessing a zero eigenvalue. To circumvent this singularity and isolate the unique physical steady-state, we replace the final linearly dependent row of $\mathbf{L}$ with the probability normalization constraint $\sum_{i=1}^{N} \rho_{ii} = 1$, yielding the modified Liouvillian matrix $\widetilde{\mathbf{L}}$. This is a typical processing method in the direct numerical solvers  \cite{tan1999computational, johansson2013qutip}, which maps the homogeneous equation into a well-posed and invertible non-homogeneous linear system:
\begin{equation}
\widetilde{\mathbf{L}} | \rho \rangle\rangle = \vec{b},
\end{equation}
where $\vec{b}$ is a column vector containing $1$ at the corresponding trace-condition index and $0$ elsewhere. 

\subsection{Macroscopic Doppler Integration}
The above steady-state vector $| \rho \rangle\rangle$ represents the response of a single atom at rest. However, the thermal vapor ensemble exhibits a Maxwell-Boltzmann velocity distribution as \cite{holloway2017electric}
\begin{equation}
p(v) = \frac{1}{\sqrt{\pi} v_{\text{th}}} e^{-\frac{v^2}{v_{\text{th}}^2}}.
\end{equation}

For an atom moving with velocity $v$ along the optical axis, 
the Hamiltonian and the resulting Liouvillian superoperator become explicitly dependent on the atomic velocity $v$ and the applied RF field $\Omega_{\text{RF}}$. For any specific velocity class $v$, the exact microscopic steady-state density vector is computationally extracted by inverting the velocity-dependent non-homogeneous linear system:
\begin{equation}
| \rho(v, \Omega_{\text{RF}}) \rangle\rangle = \widetilde{\mathbf{L}}(v, \Omega_{\text{RF}})^{-1} \vec{b}.
\end{equation}

The microscopic quantum coherence corresponding to the probe transition $\rho_{21}(v, \Omega_{\text{RF}})$ is then isolated by projecting the state vector onto the appropriate Liouville space basis element. The macroscopic atomic coherence $\bar{\rho}_{21}$ is formulated as the weighted integral over the thermal ensemble \cite{berman2011principles, ref_prajapati_sensitivity, holloway2017electric}:
\begin{equation}
\bar{\rho}_{21}(\Omega_{\text{RF}}) = \int_{-3v_{\text{th}}}^{3v_{\text{th}}} \rho_{21}(v, \Omega_{\text{RF}}) p(v) dv.
\label{ave}
\end{equation}


\begin{figure}[t]
	\centerline{\includegraphics[width=2.8in]{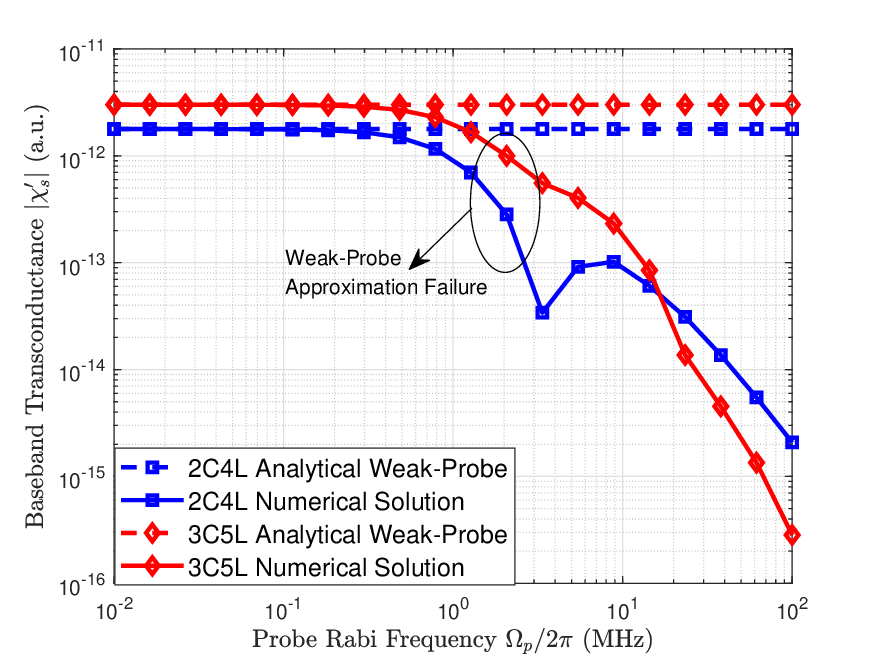}}
 \caption{ {Weak-probe analytical solution vs. numerical solution for RAQRs.}}
\label{weak}
\end{figure}

{Fig. \ref{weak} illustrates the baseband transconductance $ \chi'\subrm{s}$ as a function of the probe Rabi frequency $\Omega_p$ between the analytical weak-probe solution and the exact numerical solution derived via the Liouvillian superoperator formalism. At low probe intensities, both models exhibit a plateau characteristic of the linear response regime. The analytical model relies on a simplified average-velocity approximation, whereas the numerical solver performs a rigorous integration over the entire Maxwell-Boltzmann velocity distribution, accounting for velocity-dependent transit-time effects. As $\Omega_p$ increases, the numerical solutions for both 2C4L- and 3C5L-RAQR architectures show a significant roll-off, indicating a breakdown of the weak-probe approximation. This roll-off is attributed to atomic saturation and ground-state depletion, effects that are accurately captured by the numerical solver but ignored by the analytical model.}

\subsection{Baseband Bandwidth Characterization}
\label{ssec:transient_bandwidth}

{We characterize the dynamic response of an RAQR using two complementary bandwidth metrics. The equivalent relaxation bandwidth describes the intrinsic atomic recovery after the RF drive is removed, whereas the small-signal modulation bandwidth characterizes the actual response of a superheterodyne receiver around its LO-biased operating point.}

\subsubsection{Equivalent Transient Relaxation Bandwidth}
\label{sssec:relaxation_bandwidth}
The equivalent transient bandwidth is obtained from the free induction decay after the total RF drive is switched off \cite{meyer2018digital}. 
%
For each velocity class $v$, the initial state is defined by the exact RF-driven steady state $| \rho(v, t=0) \rangle\rangle = \widetilde{\mathbf{L}}(v, \Omega_{\text{RF}})^{-1} \vec{b}$. When the RF field is turned off at $t > 0$, the transient evolution of the density vector is governed by the homogeneous differential equation:
\begin{equation}
\frac{d}{dt} | \rho(v, t) \rangle\rangle = \mathbf{L}_{\text{off}}(v) | \rho(v, t) \rangle\rangle,
\end{equation}
where $\mathbf{L}_{\text{off}}(v)$ is the velocity-dependent Liouvillian superoperator evaluated at $\Omega_{\text{RF}} = 0$.

Rather than employing computationally expensive and unstable numerical integration methods, we obtain the exact continuous-time solution via eigendecomposition of the non-Hermitian Liouvillian matrix. Let $\mathbf{L}_{\text{off}}(v) = \mathbf{V} \mathbf{D} \mathbf{V}^{-1}$, where $\mathbf{V}$ is the matrix of right eigenvectors and $\mathbf{D}$ is the diagonal matrix of eigenvalues. The exact time-domain evolution for each velocity class is given by the matrix exponential \cite{moler2003nineteen}:
\begin{equation}
| \rho(v, t) \rangle\rangle = \mathbf{V} e^{\mathbf{D} t} \mathbf{V}^{-1} | \rho(v, 0) \rangle\rangle.
\end{equation}

Then, the macroscopic probe coherence $\bar{\rho}_{21}(t)$ is obtained by averaging over the Maxwell-Boltzmann velocity distribution $p(v)$.
Let $\bar{\rho}_{21}^{\mathrm{on}}=\bar{\rho}_{21}(0)$ and $\bar{\rho}_{21}^{\mathrm{off}}=\bar{\rho}_{21}(t\rightarrow\infty)$. The effective relaxation time $\tau_f$ is defined by the $1/e$ threshold:
\begin{equation}
\begin{split}
    \Im[\bar{\rho}_{21}(\tau_f)]
    =
    \Im[\bar{\rho}_{21}^{\mathrm{off}}]
    +
    \frac{
    \Im[\bar{\rho}_{21}^{\mathrm{on}}]
    -
    \Im[\bar{\rho}_{21}^{\mathrm{off}}]
    }{e}.
\end{split}
\label{eq:tau_f_implicit}
\end{equation}
The corresponding equivalent transient bandwidth is given by
\begin{equation}
    B_{\mathrm{rel}}
    =
    \frac{1}{2\pi\tau_f}.
\label{eq:relaxation_bandwidth}
\end{equation}
{Here, $B_{\mathrm{rel}}$ is an intrinsic recovery-bandwidth proxy rather than the modulation bandwidth of superheterodyne RAQRs.}

\subsubsection{{LO-Biased Small-Signal Modulation Bandwidth}}
\label{sssec:modulation_bandwidth}
{For superheterodyne RAQRs, the RF LO remains active during communications, where the bandwidth should be evaluated around the steady state established by the RF LO. Following the linear-response treatment of open quantum systems \cite{ban2017linear}, we define the real-valued signal perturbation as $\Omega_{\mathrm{sig}}(t) = \Re\left\{ \widetilde{\Omega}_x e^{j2\pi f_\delta t} \right\}$, where $\widetilde{\Omega}_x = \Omega_x e^{j\theta_\delta}$ is the complex amplitude of the weak RF signal. For each velocity class, we define
\begin{equation}
    \mathbf{L}_0(v) = \mathbf{L}(v,\Omega_{\mathrm{LO}}),
    \qquad
    \mathbf{L}_{\Omega}(v) = \left. \frac{\partial\mathbf{L}}{\partial\Omega_{\mathrm{RF}}} \right|_{\Omega_{\mathrm{LO}}},
\end{equation}
where $\mathbf{L}_0(v)$ is the LO-biased Liouvillian and $\mathbf{L}_{\Omega}(v)$ describes its response to the weak RF perturbation. The LO-biased steady state satisfies
\begin{equation}
    \mathbf{L}_0(v) \left|\rho_0(v)\right\rangle\rangle = 0,
    \qquad
    \operatorname{Tr}\!\left[\rho_0(v)\right] = 1.
\end{equation}
By retaining only the first-order perturbation $|\rho(v,t)\rangle\rangle = |\rho_0(v)\rangle\rangle + \Re\left\{ |\delta\rho(v,f_\delta)\rangle\rangle e^{j2\pi f_\delta t} \right\}$, the linearized frequency-domain response is given by
\begin{equation}
\begin{split}
    \left[ j2\pi f_\delta\mathbf{I} - \mathbf{L}_0(v) \right] |\delta\rho(v,f_\delta)\rangle\rangle
    =
    \mathbf{L}_{\Omega}(v) |\rho_0(v)\rangle\rangle \widetilde{\Omega}_x .
\end{split}
\label{eq:linearized_liouvillian_response}
\end{equation}
Let $\mathbf{c}^{\mathrm{T}}$ be the readout vector that extracts the detected probe-coherence quadrature. After averaging over the atomic velocity distribution, the small-signal transfer function is given by
\begin{equation}
\begin{split}
    H(f) 
    = 
    \int_{-3v_{\text{th}}}^{3v_{\text{th}}} p(v)\, \mathbf{c}^{\mathrm{T}} \left[ j2\pi f\mathbf{I} - \mathbf{L}_0(v) \right]^{-1} \mathbf{L}_{\Omega}(v) \left|\rho_0(v)\right\rangle\rangle \,dv .
\end{split}
\label{eq:raqr_baseband_transfer}
\end{equation}
Here, the inverse in Eq. \eqref{eq:raqr_baseband_transfer} is evaluated on the trace-zero perturbation subspace.
The normalized baseband response is given by $\mathcal{H}(f)=\frac{H(f)}{H(0)}$. The LO-biased small-signal modulation bandwidth is defined by the first $-3$ dB crossing connected to the zero-frequency passband:
\begin{equation}
    B_{\mathrm{mod}} = \inf\left\{ f>0: \left| \mathcal{H}(f) \right|^2 \leq \frac{1}{2} \right\}.
\label{eq:modulation_bandwidth}
\end{equation}
When the response is dominated by a single relaxation pole and the RF LO does not significantly modify the atomic dynamics, we have $B_{\mathrm{mod}} \approx B_{\mathrm{rel}}$.}

\section{Numerical Results}

\begin{table}[t]
\centering
\caption{Atomic Parameters for {2C4L- and 3C5L-RAQRs}}
\label{tab:architecture_params_completed}
\begin{tabular}{l l l}
\hline
\textbf{Parameters} & {\textbf{2C4L-RAQR }} & {\textbf{3C5L-RAQR}} \\
\hline
Vapor Cell Length $d$ &7.5 cm & $7.5$ cm \\
Atom Density $N_a$ & $1.5\times10^{11}$$\text{cm}^{-3}$ & $1.5\times10^{11}$$\text{cm}^{-3}$\\
Probe Laser $\lambda_\text{p}$ & 852 nm & 895 nm \\
Dressing Laser $\lambda_\text{d}$ & N/A & 636 nm \\
Coupling Laser $\lambda_\text{c}$ & 510 nm  & 2245 nm \\
Probe Beam Power $\mathcal{P}_0$ & $1$ mW & $50$ $\mu$W  \\
Coupling Beam Power $\mathcal{P}_\text{c}$ & $150 $ mW  & $40$ mW  \\
Dressing Beam Power $\mathcal{P}_\text{d}$ & N/A & $20$ mW  \\
Beam Radius $r_0$ & $0.38$ mm  & $0.38$ mm  \\
RF LO Field $U_y$ & $0.03$ V/m  & $0.03$ V/m  \\
Dipole Moment $\mu_{12}$ & $2.59\text{ }q_0a_0$  & $1.84$ $q_0a_0$  \\
Dipole Moment $\mu_{23}$ & $0.022\text{ }q_0a_0$  & $0.23$ $q_0a_0$  \\
Dipole Moment $\mu_{34}$ & $1443.48\text{ }q_0a_0$  & $0.019$ $q_0a_0$  \\
Dipole Moment $\mu_{45}$ & N/A & $1443.48$ $q_0a_0$ \\
Natural Decay Rates $\Gamma_2 /2\pi $ &  5.22 MHz  & $ 4.56$ MHz  \\
Natural Decay Rates $\Gamma_3 /2\pi $ &  2.73 kHz  & $ 0.98$ MHz  \\
Natural Decay Rates $\Gamma_4 /2\pi $ & 0.57 kHz  & $ 0.57$ kHz  \\
Natural Decay Rates $\Gamma_{5}/2\pi $ & N/A & $ 2.73$ kHz  \\
Residual Dephasing Rates $\Gamma_{\text{d}}/2\pi$ & 50 kHz & $ 50$ kHz  \\
\hline
\end{tabular}
\end{table}
\label{sec:numerical_results}



Unless otherwise specified, we set $G_\text{t}=10$ dB, $G_\text{r}=5.5$ dB, $G_\text{LNA}=60$ dB, and $F=6$ for the classical RF receiver, while the photoelectric detection parameters of RAQRs are set to $G_{\text{PD}}=30$ dB, $T_\text{atom}=T_{\text{PD}}=T_{\text{env}}=290$ K,  $\alpha=0.8$ and $\mathcal{P}_l=30$ mW \cite{ref_gong_overview}. The distance $L$ between the transmitter and the receivers is set to $L=10$ m. Table \ref{tab:architecture_params_completed} presents the specific atomic parameters for 2C4L- and 3C5L-RAQRs. The intrinsic properties of Rydberg atoms are computed by utilizing the open-source ARC library and publicly available measured data \cite{vsibalic2017arc, steck2025rubidium}, e.g., dipole moments $\mu_{ij}$ and natural decay rates $\Gamma_{n}$, while several controllable parameters of RAQRs, e.g., RF LO field $U_y$ and laser powers, are systematically optimized to maximize the system-level SNR for each architecture. {In particular, the different laser powers used for the 2C4L- and 3C5L-RAQRs correspond to independently optimized SNR operating points under the physical constraints. For the 2C4L-RAQR, increasing the probe power partially compensates for the reduced atomic transconductance caused by residual Doppler broadening. In contrast, the smaller residual Doppler mismatch of the 3C5L-RAQR enables its SNR-oriented operating point to be reached at a lower probe power.}

%

\begin{figure}[t]
	\centerline{\includegraphics[width=2.8in]{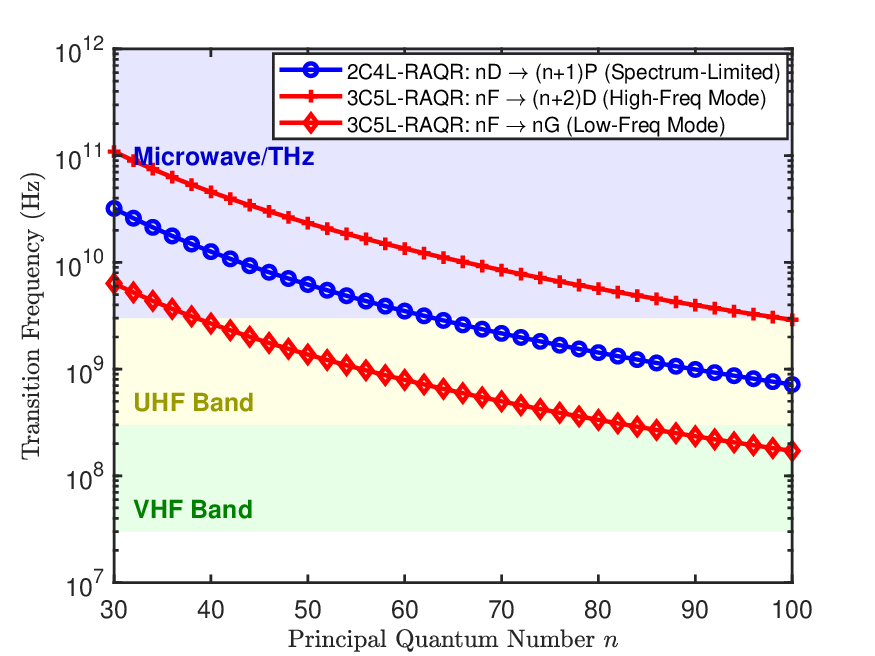}}
\caption{ Spectrum access range of RAQRs.}
\label{fig:access}
\end{figure}

\begin{figure}[t]
	\centerline{\includegraphics[width=2.8in]{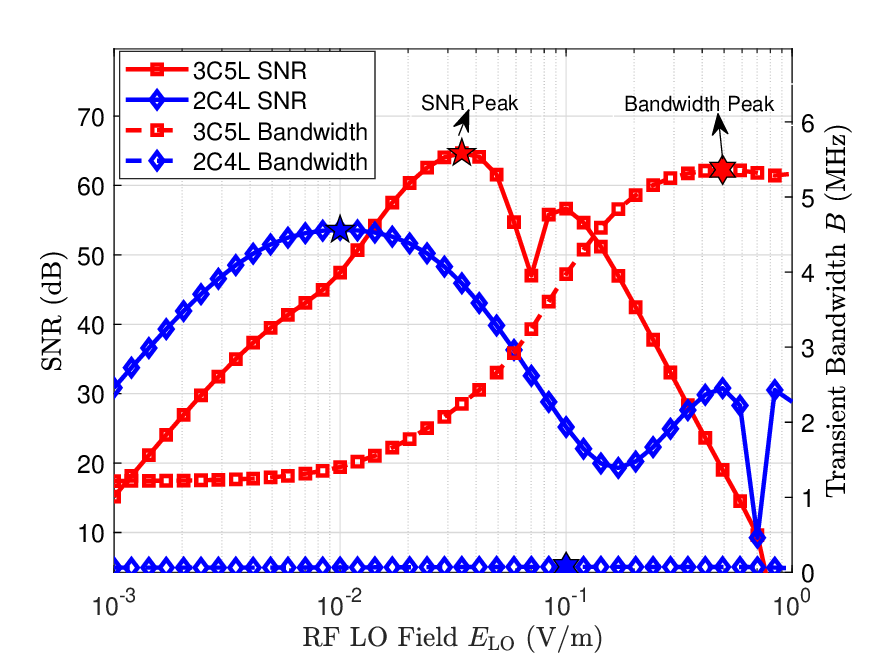}}
 \caption{SNR-bandwidth trade-off of RAQRs.}
\label{fig:tradeoff}
\end{figure}

\begin{figure}[t]
	\centerline{\includegraphics[width=2.8in]{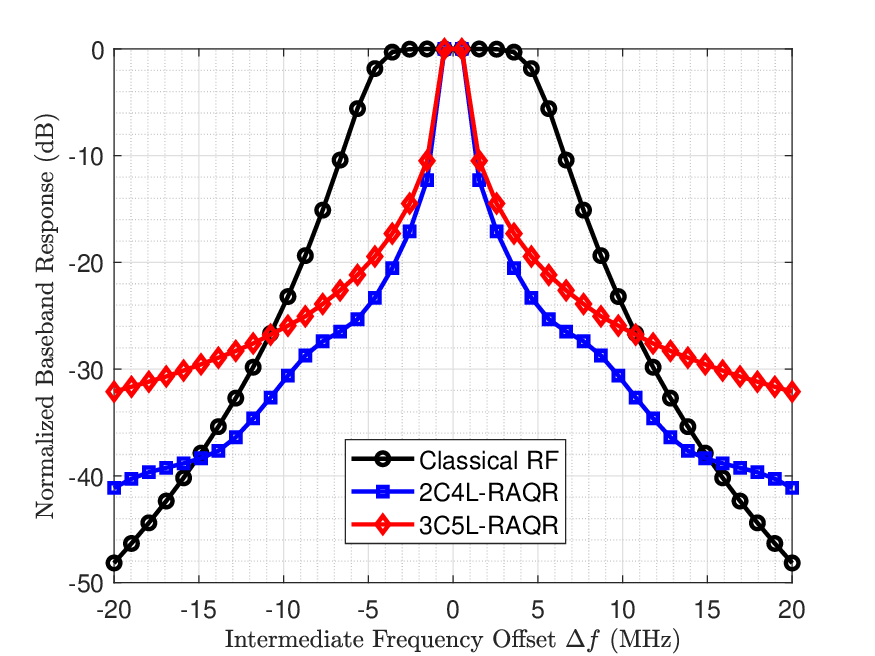}}
\caption{Frequency selectivity profiles for different receivers.}
\label{fig:filter}
\end{figure}

Fig. \ref{fig:access} presents the accessible RF frequencies against the principal quantum number $n$ for different setups of RAQRs. 
While the 2C4L-RAQR can theoretically access the upper UHF band by employing very high principal quantum numbers, it is physically constrained by the large energy gaps inherent in the $P \to D$ pathway. 
In contrast, the 3C5L-RAQR provides more flexible access to the UHF/VHF bands by utilizing high-angular-momentum transitions, which naturally reside in lower frequency ranges even at moderate $n$ values. {Since the dipole moment $\mu$ generally scales with $n^2$ \cite{gallagher1994rydberg, fancher2021rydberg}, the 2C4L-RAQR requires excessively high principal quantum numbers to reach the UHF/VHF band. This leads to a fragile sensitivity where the massive dipole moment makes the system extremely vulnerable to environmental noise and background fluctuations.} 
To fairly compare the performance between 2C4L- and 3C5L-RAQRs, both RAQRs receive the impinging signal with the same carrier frequency $f_c$ in this work. We consider an RF signal-induced atomic transition $\overleftrightarrow {\ket{47D_{5/2}}, \ket{48P_{3/2}}}$ as illustrated in Fig. \ref{fig:level_schemes}, and hence $f_c = 6.94 $ GHz according to Eq. \eqref{fre}. 

Fig. \ref{fig:tradeoff} unveils the fundamental physical trade-off between the system sensitivity and the instantaneous bandwidth $B_\text{RAQR}$ for RAQRs, modulated by the RF LO field $U_y$. Operating at a low $U_y$ minimizes the power broadening effect, thereby preserving an ultra-sharp susceptibility peak $\chi_{\text{s}}^\prime$ that maximizes the SNR. However, this sharp resonance dictates a long-lived coherent state, resulting in a sluggish transient relaxation time $\tau_f$ and a collapsed communication bandwidth. Conversely, increasing $U_y$ accelerates the Rabi oscillations and forces rapid state evolution, which significantly expands the bandwidth but flattens the atomic dispersion curve, severely degrading the SNR. 
{Consequently, it is physically impossible to achieve optimal bandwidth and SNR simultaneously using a static set of parameters. In practical deployments, the RAQR physical parameters must be dynamically configured. Specifically, the SNR-centric optimization is mandatory for ultra-weak signal environments, e.g., edge-of-cell coverage or deep-space telemetry, where closing the link budget is the primary bottleneck, whereas bandwidth-centric optimization is required for capacity-driven links operating well above the fundamental noise floor.} In the performance evaluation of this work, we adopt SNR-centric optimization to determine the physical parameters of the RAQR.


Fig. \ref{fig:filter} compares the frequency selectivity profiles for different receiver architectures. Specifically, the normalized baseband response for RAQRs is obtained by computing the magnitude of the atomic transconductance $|\chi'_\text{s}(\Delta f)|$, while the response for the classical RF receiver is modeled using a standard Butterworth filter with bandwidth $B_\text{CL}$=10 MHz \cite{pfister2017discrete}. Although the classical RF receiver exhibits a broader and flatter response, the RAQRs demonstrate a significantly sharper selectivity profile. 
{The sluggish transient response of the RAQR provides inherent robustness against hardware impairments such as oscillator phase noise, where any high-frequency phase jitter originating from the local RF oscillators is heavily attenuated by the narrow low-pass filtering effect of the atomic ensemble. }


\begin{figure}[t]
	\centerline{\includegraphics[width=2.8in]{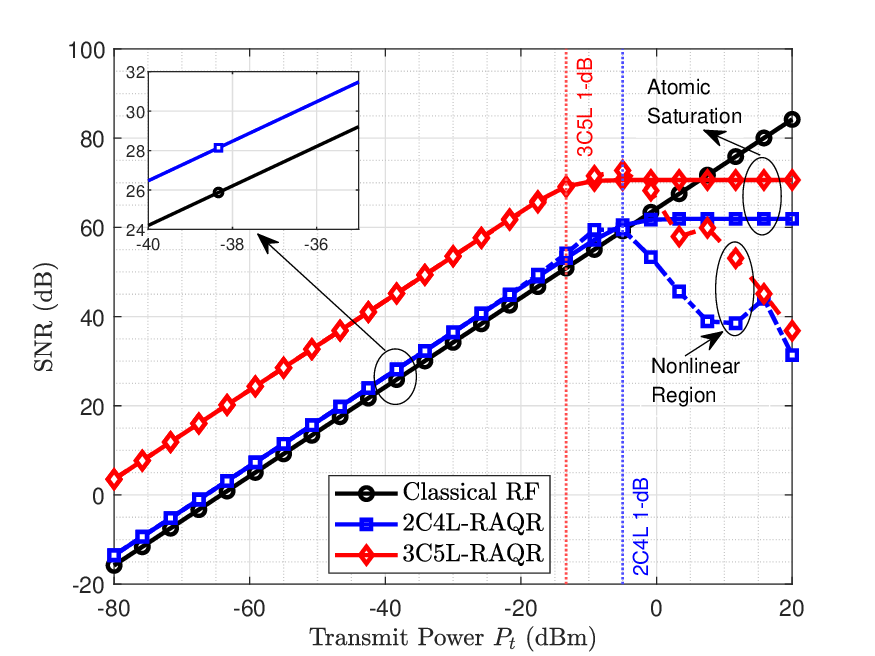}}
\caption{Transmit power $P_t$ vs. SNR for different receivers.}
\label{fig:SNR}
\end{figure}

\begin{figure}[t]
	\centerline{\includegraphics[width=2.8in]{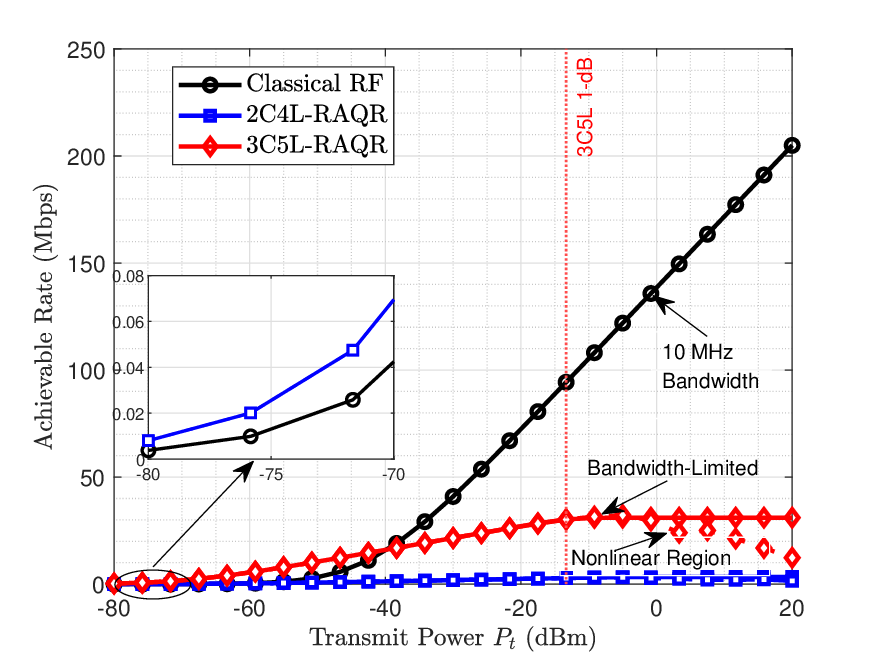}}
 \caption{ {Transmit power $P_t$ vs. achievable rate for different receivers.}}
\label{fig:Rate}
\end{figure}

\begin{figure}[t]
	\centerline{\includegraphics[width=2.8in]{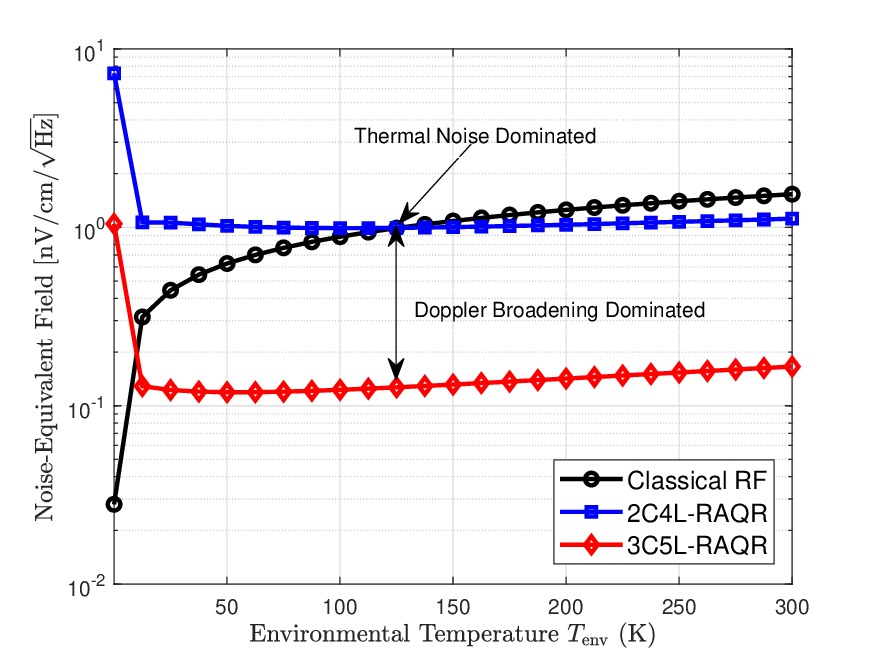}}
\caption{ {Noise-equivalent field vs. environmental temperature $T_{\mathrm{env}}$.}}
\label{Tep}
\end{figure}

\begin{figure}[t]
	\centerline{\includegraphics[width=2.8in]{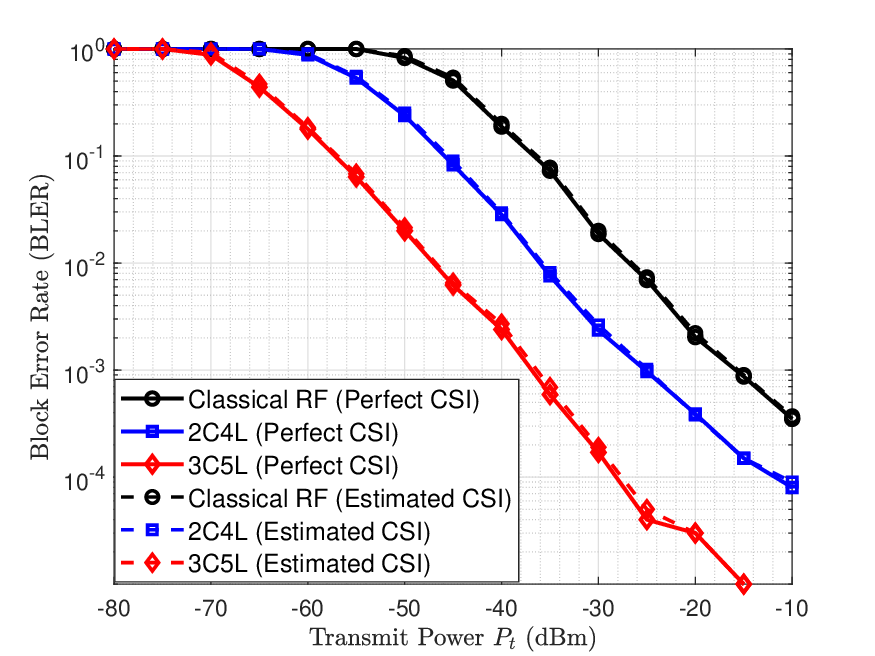}}
\caption{{ Transmit power $P_t$ vs. BLER for different receivers.}}
\label{fig:BLER}
\end{figure}

{Fig. \ref{fig:SNR} presents the received SNR against the transmit power $P_t$ for different receivers, where a common 100-kHz post-detection noise bandwidth is used for a fair SNR comparison. Thanks to the suppression of residual Doppler broadening, the 3C5L-RAQR significantly outperforms the 2C4L-RAQR, while the SNR performance between 2C4L-RAQR and classical RF receiver is close. As the incident signal power increases and approaches the LO field strength, the superheterodyne approximation is no longer valid. In this work, we identify the first 1-dB compression point that is defined as the lowest input power at which the fixed-LO fundamental conversion gain falls 1 dB below its small-signal linear extrapolation.
Beyond the 1-dB compression point, Fig.~\ref{fig:SNR} presents two numerical treatments. The first is the degraded nonlinear response obtained from the fundamental Liouvillian solution. The second, referred to as the managed atomic-saturation model, assumes practical dynamic-range control, e.g., gain control or switching to a strong-field detection mode, and represents the post-compression performance by a smooth saturation-limited plateau calibrated at the 1-dB compression point. Since the classical RF receiver utilizes sophisticated automatic gain control amplifiers to handle signal swings, it remains superior in scenarios characterized by large dynamic power range variations, while the RAQR is a superior choice for typical weak-signal detection scenarios, e.g., long-range communications and low-power networks, where link budget is the primary constraint.}

{Fig. \ref{fig:Rate} compares the achievable {rate} for different receivers, where the feasible bandwidths $B_\text{RAQR}$ of RAQRs are determined by Section IV-C and hence $B_\text{CL} = 10$ MHz $ >B_\text{RAQR} $.} In the transmitted power-limited regime, i.e., low received SNR $\gamma$, approximating $\log(1+\gamma) \approx \gamma/\ln2$, the {achievable rate} scales as $C \approx B  \gamma/\ln2$. The high SNR advantages of RAQRs can provide a higher achievable {rate} compared to the classical RF receiver. However, in the high transmitted power regime, approximating $\log(1+\gamma) \approx \log(\gamma)$, the {achievable rate} scales as $C \approx B \log(\gamma)$. The linear reduction in bandwidth $B$ dominates the logarithmic gain in SNR for RAQRs. Thus, for high-power and high-speed links, the classical RF receiver remains superior. In particular, the 3C5L-RAQR exhibits a higher baseline transient bandwidth compared to the 2C4L-RAQR, and hence provides a higher achievable rate.

{Fig. \ref{Tep} presents the noise-equivalent field (NEF) to characterize the receiver sensitivity with respect to the environmental temperature $T_{\text{env}}$ \cite{ref_gong_overview}, where the reference incident signal field strength is set to 1 $ \mathrm{ nV}/\mathrm{cm}$. 
The classical RF antenna performs exceptionally well at cryogenic temperatures but its sensitivity degrades rapidly as it becomes dominated by Johnson-Nyquist thermal noise at higher temperatures. 
Conversely, the 2C4L-RAQR surpasses the classical antenna as $T_{\text{env}}$ increases, since its optical readout mechanism is largely immune to direct electronic thermal noise. However, the NEF of the 2C4L-RAQR still gradually deteriorates because the increased thermal velocity of the atoms exacerbates the residual Doppler broadening. Here, the atom density $N_a$ within the vapor cell during the temperature sweep is assumed constant to isolate the explicit temperature dependence of the velocity distribution and noise terms, while the actual atom density will vary with the temperature of a vapor cell \cite{ref_gong_overview}.
By effectively cancelling the Doppler mismatch through its three-photon resonance, the 3C5L-RAQR maintains a highly robust and superior sensitivity over the simulated temperature range under the fixed-density assumption. Moreover, when $T_{\text{env}}$ approaches zero for RAQRs, the thermal velocity of the quasi-cold atoms $v_{\text{th}}$ becomes infinitesimally small, causing the transit-time relaxation rate $v_{\text{th}} / r_{\text{0}}$ to approach zero. Under continuous-wave laser illumination, these near-stationary atoms are persistently excited, leading to severe optical pumping and ground-state depletion. Without the continuous thermal replenishment provided by fresh atoms traversing the laser beams, the macroscopic atomic polarization collapses, manifesting as the dramatic NEF spike observed at the low-temperature boundary of adopted fixed-density continuous-wave model. In this case, to operate Rydberg electrometers effectively in ultra-low-temperature regimes, continuous-wave interrogation needs to be replaced by pulsed excitation sequences, e.g., Ramsey interferometry \cite{facon2016sensitive}, to preserve the coherent atomic response\footnote{{It is crucial to reconcile the macroscopic sensitivity degradation at ultra-low temperatures with the idealized spectral lineshape previously presented in Fig. \ref{fig:4L_spectrum2} of Section II-A. Fig. \ref{fig:4L_spectrum2} illustrates the normalized steady-state absorption spectrum $\Im(\hat{\rho}_{21})$, highlighting the pure kinematic advantage of cryogenic temperatures by suppressing the Doppler broadening. However, optimal spectral resolution does not linearly translate to optimal continuous-wave communication performance. The system-level NEF metric captures the dynamic macroscopic response, which relies on the continuous replenishment of ground-state atoms, resulting in sensitivity collapse at ultra-low $T_\text{env}$.}}.
Note that the state-of-the-art practical experiments have pushed RAQR sensitivities to the scale of $\text{nV/cm}/\sqrt{\text{Hz}}$ regime \cite{ref_jing_superheterodyne, yuan2023quantum}, while these implementations remain inherently constrained by unavoidable macroscopic technical noise. In this work, we isolate unpredictable engineering bottlenecks in the practical deployment to derive the theoretical NEF boundary.}

{Fig.~\ref{fig:BLER} presents the BLER performance obtained via a link-level simulation, where the number of transmitted data blocks is $1\times10^4$ and each block consists of 512 symbols. Note that we evaluate the BLER rather than the BER because practical communication systems utilizing forward error correction process and decode data in discrete blocks. Consequently, since any uncorrectable bit error compromises the entire block, BLER serves as a more accurate metric for determining actual link reliability and effective throughput.}
The system employs 16-QAM modulation over a Rayleigh fading channel, utilizing a least-squares estimator to acquire the channel state information (CSI) and a linear minimum mean-square error detector for signal recovery.
We observe that the superior atomic sensitivity of the 3C5L-RAQR translates into a large BLER gain.

\section{Conclusion}
\label{sec:conclusion}
In this paper, we investigated the 3C5L-RAQR architecture and exhibited an equivalent baseband signal model for wireless communications, which effectively alleviates the residual Doppler broadening in thermal vapor cells. We provided a systematic performance analysis and the exact numerical solution method for RAQRs. Numerical analyses reveal that the 3C5L-RAQR achieves superior sensitivity, particularly in power-limited regimes and can serve as a versatile unified platform, capable of accessing the strategic UHF/VHF spectrum via high-angular-momentum transitions.
We also highlighted a no-free-lunch trade-off in RAQR design, i.e., the extreme sensitivity of the RAQR comes at the cost of a narrower instantaneous bandwidth and a lower saturation threshold. 
{Moreover, while current RAQR prototypes rely on discrete optical components and laser locking electronics, indicating a substantially higher integration complexity than single-chip integrated RF systems, their scalability trajectory is firmly rooted in the rapid advancement of photonic integrated circuits and microelectromechanical-system (MEMS) vapor cells. Future research will focus on enhancing the communication rate of the 3C5L-RAQR by integrating frequency-division or spatial multiplexing techniques \cite{gong2025rydberg}, and the compact RAQR deployment in standalone communication and sensing platforms.}

\appendices
\section{Derivation of 3C5L Atomic Coherence ${\rho}_{21}^{{\text{5L}}}$}
\label{app:derivation_5level}
To obtain an analytical form of the steady-state coherence $\rho_{21}^{\rm{5L}}$, we first determine the commutator $[\hat{H}_{\rm{5L}}, \hat{\rho}]_{n1}$ based on the master equation $d\hat{\rho}/dt = 0$. For the $5 \times 5$ complex Hamiltonian matrix $\hat{H}_{\rm{5L}}$ defined in Eq. \eqref{eq:rx_hamiltonian_5level}, the commutator is given by $[\hat{H}_{\rm{5L}}, \hat{\rho}]_{ij} = \sum_{k=1}^5 H_{ik}\rho_{kj} - \rho_{ik}H_{kj}$. Thus, for the probe transition coherence $[\hat{H}_{\rm{5L}}, \hat{\rho}]_{21}$, we formulate it as
\begin{equation}
[\hat{H}_{\rm{5L}}, \hat{\rho}]_{21} = \frac{\hbar}{2}(\Omega_p^{\rm{5L}})^*(\rho_{11} - \rho_{22}) - \hbar\Delta_2^{\rm{5L}}\rho_{21} + \frac{\hbar}{2}\Omega_d^{\rm{5L}}\rho_{31}.
\end{equation}
According to the Lindblad master equation, the time evolution of $\rho_{21}$ is given by
\begin{equation}
\dot{\rho}_{21} = -\frac{i}{2}(\Omega_p^{\rm{5L}})^*(\rho_{11} - \rho_{22}) + i\Delta_2^{\rm{5L}}\rho_{21} - \frac{i}{2}\Omega_d^{\rm{5L}}\rho_{31} - \Gamma_{21}\rho_{21}.
\end{equation}
Similarly, we obtain the evolution equations for the higher-order coherences $\rho_{31}$, $\rho_{41}$, and $\rho_{51}$. For notational conciseness, we omit the superscript 5L in the following derivation steps:
\begin{equation}
\dot{\rho}_{31} = -\frac{i}{2}\Omega_d^*\rho_{21} + i\Delta_3\rho_{31} - \frac{i}{2}\Omega_c\rho_{41} + \frac{i}{2}\Omega_p^*\rho_{32} - \Gamma_{31}\rho_{31},
\end{equation}
\begin{equation}
\dot{\rho}_{41} = -\frac{i}{2}\Omega_c^*\rho_{31} + i\Delta_4\rho_{41} - \frac{i}{2}\Omega_{\rm{RF}}\rho_{51} + \frac{i}{2}\Omega_p^*\rho_{42} - \Gamma_{41}\rho_{41},
\end{equation}
\begin{equation}
\dot{\rho}_{51} = -\frac{i}{2}\Omega_{\rm{RF}}^*\rho_{41} + i\Delta_5\rho_{51} + \frac{i}{2}\Omega_p^*\rho_{52} - \Gamma_{51}\rho_{51}.
\end{equation}
The steady-state condition enforces $\dot{\rho}_{21} = \dot{\rho}_{31} = \dot{\rho}_{41} = \dot{\rho}_{51} = 0$. By applying the weak-probe approximations, i.e., $\rho_{11} \approx 1$, $\rho_{ii \neq 1} \to 0$, and $\rho_{j2} \to 0$, the coupled system can be recast as
\begin{equation}
(i\Delta_2 - \Gamma_{21})\rho_{21} - \frac{i}{2}\Omega_d\rho_{31} = \frac{i}{2}\Omega_p^*, \label{eq:a6}
\end{equation}
\begin{equation}
(i\Delta_3 - \Gamma_{31})\rho_{31} - \frac{i}{2}\Omega_d^*\rho_{21} - \frac{i}{2}\Omega_c\rho_{41} = 0, \label{eq:a7}
\end{equation}
\begin{equation}
(i\Delta_4 - \Gamma_{41})\rho_{41} - \frac{i}{2}\Omega_c^*\rho_{31} - \frac{i}{2}\Omega_{\rm{RF}}\rho_{51} = 0, \label{eq:a8}
\end{equation}
\begin{equation}
(i\Delta_5 - \Gamma_{51})\rho_{51} - \frac{i}{2}\Omega_{\rm{RF}}^*\rho_{41} = 0. \label{eq:a9}
\end{equation}

To isolate $\rho_{21}$, we iteratively back-substitute the higher-order coherences. From Eq. (\ref{eq:a9}), we express $\rho_{51}$ as
\begin{equation}
\rho_{51} = \frac{i\Omega_{\rm{RF}}^*/2}{(i\Delta_5 - \Gamma_{51})}\rho_{41}.
\end{equation}
Substituting $\rho_{51}$ into Eq. (\ref{eq:a8}), we obtain:
\begin{equation}
\left[ (i\Delta_4 - \Gamma_{41}) + \frac{|\Omega_{\rm{RF}}|^2/4}{i\Delta_5 - \Gamma_{51}} \right] \rho_{41} = \frac{i}{2}\Omega_c^*\rho_{31}.
\end{equation}
By defining the nested denominator $f_5 = i\Delta_5 - \Gamma_{51}$ and $f_4 = (i\Delta_4 - \Gamma_{41}) + |\Omega_{\rm{RF}}|^2/(4f_5)$, we have $\rho_{41} = \frac{i\Omega_c^*/2}{f_4}\rho_{31}$. 
Following the same iterative procedure, we substitute $\rho_{41}$ into Eq. (\ref{eq:a7}) to obtain $\rho_{31} = \frac{i\Omega_d^*/2}{f_3}\rho_{21}$, where $f_3 = (i\Delta_3 - \Gamma_{31}) + |\Omega_c|^2/(4f_4)$. Finally, substituting $\rho_{31}$ into Eq. \eqref{eq:a6} isolates $\rho_{21}$, yielding the continued-fraction solution in Eq. \eqref{eq:rho21_5level_continued_fraction}.


\section{Derivation of the BCOD AC Voltage $\tilde{V}(t)$}
\label{app:derivation_BCOD}

The differential term $P'\subrm{mix}$ for $\Omega\subrm{RF}$ in AC voltage $\tilde{V}(t)$ can be expressed as
\begin{align}
P'\subrm{mix} = \frac{d\sqrt{\mathcal{P}_1}}{d\Omega\subrm{RF}} \cos(\phi_l - \phi\subrm{p}) + \sqrt{\mathcal{P}_1} \frac{d \cos(\phi_l - \phi\subrm{p})}{d\Omega\subrm{RF}}.
\label{eq:app_bcod1}
\end{align}
Using $\frac{d\sqrt{f}}{dx} = \frac{1}{2\sqrt{f}} \frac{df}{dx}$, we have
\begin{align}
\frac{d\sqrt{\mathcal{P}_1}}{d\Omega\subrm{RF}} &= \frac{1}{2\sqrt{\mathcal{P}_1}} \left[ \mathcal{P}_1 \left( -\frac{2\pi d}{\lambda\subrm{p}} \Im(\chi'\subrm{s}) \right) \right] =  -\frac{\pi d\sqrt{\mathcal{P}_1}}{\lambda\subrm{p}} \Im(\chi'\subrm{s}),
\end{align}
\begin{align}
\frac{d \cos(\phi_l - \phi\subrm{p})}{d\Omega\subrm{RF}} &= \sin(\phi_l - \phi\subrm{p}) \frac{d\phi\subrm{p}}{d\Omega\subrm{RF}} = \sin(\phi_l - \phi\subrm{p}) \frac{\pi d}{\lambda\subrm{p}} \Re(\chi'\subrm{s}).
\end{align}
Substituting these back into $P'\subrm{mix}$ at $\Omega\subrm{LO}$:
\begin{align}
P'\subrm{mix} = \frac{\pi d \sqrt{\mathcal{P}_1}}{\lambda\subrm{p}} \left( \Re(\chi'\subrm{s}) \sin(\phi_l - \phi\subrm{p}) - \Im(\chi'\subrm{s})\cos(\phi_l - \phi\subrm{p})\right).
\end{align}
Using $\sin(A-B) = \sin A \cos B - \cos A \sin B$ and $\chi'\subrm{s} = |\chi'\subrm{s}|
e^{i\psi\subrm{p}}$, we have
\begin{align}
&P'\subrm{mix} = \frac{\pi d \sqrt{\mathcal{P}_1}}{\lambda\subrm{p}} |\chi'\subrm{s}| ( \cos(\psi\subrm{p}) \sin(\phi_l - \phi\subrm{p})  \nonumber \\
&- \sin(\psi\subrm{p}) \cos(\phi_l - \phi\subrm{p}) )= \frac{\pi d \sqrt{\mathcal{P}_1}}{\lambda\subrm{p}} |\chi'\subrm{s}| \sin(\phi_l - \phi\subrm{p} - \psi\subrm{p}).
\end{align}
Letting $\varphi_1 = \phi_l - \phi\subrm{p} - \psi\subrm{p}$, we have
\begin{equation}
P'\subrm{mix}(\Omega\subrm{LO}) = \frac{\pi d \sqrt{\mathcal{P}_1}}{\lambda\subrm{p}} |\chi'\subrm{s}| \sin(\varphi_1).
\end{equation}
Substituting this into $\tilde{V}(t) = 2 \alpha \sqrt{{G_{\text{PD}}}\mathcal{P}_l} \Omega\subrm{sig}(t) P'\subrm{mix}(\Omega\subrm{LO})$, we arrive at Eq. \eqref{eq:passband_voltage_simple}.

\section{Derivation of Atomic Transconductance $\chi'\subrm{s}$}
\label{app:derivation_transconductance}
We decompose the derivative $\chi'\subrm{s} = \left. \frac{\partial \chi}{\partial \Omega\subrm{RF}} \right|_{\Omega\subrm{LO}}$ into three parts. First, we define the RF coupling term from the continued fractions in Appendix \ref{app:derivation_5level} as $C_{\text{RF}} = |\Omega_{\text{RF}}|^2 / 4$. The chain rule of 3C5L-RAQR is given by
\begin{equation}
\chi'\subrm{s,{{\text{5L}}}} = \frac{\partial \chi}{\partial \rho^{{\text{5L}}}_{21}}  \frac{\partial {\rho}_{21}^{{\text{5L}}} }{\partial C_{\text{RF}}}  \frac{\partial C_{\text{RF}}}{\partial \Omega_{\text{RF}}}.
\label{eq:a1}
\end{equation}
We evaluate each term in Eq. \eqref{eq:a1}:
\begin{enumerate}
    \item {$\partial \chi / \partial {\rho}_{21}^{{\text{5L}}} $:} From Eq. \eqref{eq:rx_susceptibility}, this is a constant $K$:
    \begin{equation}
    \frac{\partial \chi}{\partial {\rho}_{21}^{{\text{5L}}} } = K = \frac{2N_a |\mu_{12}|^2}{\epsilon_0 \hbar \Omega_p}.
    \end{equation}
    
    \item {$\partial C_{\text{RF}} / \partial \Omega_{\text{RF}}$:} Assuming a real-valued $\Omega_{\text{LO}}$ bias:
    \begin{equation}
    \frac{\partial C_{\text{RF}}}{\partial \Omega_{\text{RF}}} = \frac{\partial}{\partial \Omega_{\text{RF}}} \left( \frac{\Omega_{\text{RF}}^2}{4} \right) = \frac{\Omega_{\text{RF}}}{2}.
    \end{equation}
    Evaluated at the LO bias, this gives $\frac{\Omega_{\text{LO}}}{2}$.
    
    \item {$\partial {\rho}_{21}^{{\text{5L}}}  / \partial C_{\text{RF}}$:} 
We define the nested denominators in Eq. \eqref{eq:rho21_5level_continued_fraction} evaluated at $\Omega_{\text{LO}}$ as
\begin{align}
& f_5 = i\Delta_5 - \Gamma_{51}, \\
& f_4 = (i\Delta_4 - \Gamma_{41}) + |\Omega_{\text{LO}}|^2 / (4f_5), \\
& f_3 = (i\Delta_3 - \Gamma_{31}) + |\Omega\subrm{c}|^2 / (4 f_4), \\
& f_2 = (i\Delta_2 - \Gamma_{21}) + |\Omega\subrm{d}|^2 / (4 f_3).  
\end{align}
We apply the chain rule $\frac{\partial {\rho}_{21}^{{\text{5L}}} }{\partial C_{\text{RF}}} = \frac{\partial {\rho}_{21}^{{\text{5L}}} }{\partial f_2} \cdot \frac{\partial f_2}{\partial f_3} \cdot \frac{\partial f_3}{\partial f_4} \cdot \frac{\partial f_4}{\partial C_{\text{RF}}}$:
\begin{align}
\frac{\partial{\rho}_{21}^{{\text{5L}}} }{\partial C_{\text{RF}}} &=  \frac{-i\Omega\subrm{p}/2}{f_2^2} \frac{-|\Omega\subrm{d}|^2/4}{f_3^2}  \frac{-|\Omega\subrm{c}|^2/4}{f_4^2} \frac{1}{f_5}.
\label{eq:deriv_rho21_5L}
\end{align}
\end{enumerate}

Hence, the full 3C5L atomic transconductance is given by
\begin{equation}
\chi'\subrm{s, \text{5L}} = \frac{K \Omega_{\text{LO}}}{2}  \left[ \frac{ -(i\Omega\subrm{p}/2) (|\Omega\subrm{d}|^2/4) (|\Omega\subrm{c}|^2/4) }{ f_2^2 f_3^2 f_4^2 f_5 } \right].
\end{equation}

The 2C4L derivation is a subset of the above, where $f_2 = (i\Delta_2 - \Gamma_{21}) + |\Omega\subrm{c}|^2 / (4 f_3)$ and $f_3 = (i\Delta_3 - \Gamma_{31}) + C_{\text{RF}} / f_4$. The chain rule is shorter, which is given by
\begin{align}
\frac{\partial \rho_{21}^\text{4L}}{\partial C_{\text{RF}}} &= \frac{\partial \rho_{21}}{\partial f_2}  \frac{\partial f_2}{\partial f_3}  \frac{\partial f_3}{\partial C_{\text{RF}}}= \frac{-i\Omega\subrm{p}/2}{f_2^2}\frac{-|\Omega\subrm{c}|^2/4}{f_3^2}\frac{1}{f_4}.
\label{eq:deriv_rho21_4L}
\end{align}

The full 2C4L atomic transconductance is given by
\begin{equation}
\chi'\subrm{s, \text{4L}} = \frac{K \Omega_{\text{LO}}}{2} \frac{ (i\Omega\subrm{p}/2) (|\Omega\subrm{c}|^2/4) }{ f_2^2 f_3^2  f_4 }\label{{eq:chi_derivative_appendix}}.
\end{equation}

\bibliographystyle{IEEEtran}
\bibliography{IEEEabrv,refs_5LRAQR.bib}

\end{document}